\pdfoutput=1 
\documentclass{JINST}
\usepackage[caption=false]{subfig}
\usepackage{mathtools}
\usepackage[]{placeins}
\usepackage{multirow}
\usepackage[]{threeparttable}

\title{First Measurements with New High-Resolution Gadolinium-GEM Neutron Detectors}

\author{
Dorothea~Pfeiffer$^{a,b}$\thanks{Corresponding
author.}$\ $,
Filippo~Resnati$^{a,b}$,
Jens~Birch$^{c}$,
Maddi~Etxegarai$^{a}$,
Richard~Hall-Wilton$^{a,d}$, 
Carina~H\"{o}glund$^{a,c}$,
Lars~Hultman$^{d}$,
Isabel~Llamas-Jansa$^{a,e}$,
Eraldo~Oliveri$^{b}$,
Esko~Oksanen$^{a}$,
Linda~Robinson$^{a}$,
Leszek~Ropelewski$^{b}$,
Susann~Schmidt$^{a,c}$,
Christina~Streli$^{f}$,
Patrik Thuiner$^{b,f}$
\\
\llap{$^a$}European Spallation Source (ESS AB)\\
	P.O. Box 176, SE-22100 Lund, Sweden\\
\llap{$^b$}CERN\\
  	CH-1211 Geneva 23, Switzerland\\
\llap{$^c$}Link\"{o}ping University\\
	IFM SE-581 83 Link\"{o}ping, Sweden\\
\llap{$^d$}Mid-Sweden University\\
	SE-85170 Sundsvall, Sweden\\
\llap{$^e$}Institute for Energy Technology IFE\\
	NO-2007 Kjeller, Norway\\	
\llap{$^f$}Vienna University of Technology\\
	1040 Vienna, Austria\\

E-mail: \email{Dorothea.Pfeiffer@cern.ch}}

\abstract{European Spallation Source instruments like the macromolecular diffractometer (NMX) require an excellent neutron detection efficiency, high-rate capabilities, time resolution, and an unprecedented spatial resolution in the order of a few hundred micrometers over a wide angular range of the incoming neutrons. For these instruments solid converters in combination with Micro Pattern Gaseous Detectors (MPGDs) are a promising option. A GEM detector with gadolinium converter was tested on a cold neutron beam at the IFE research reactor in Norway. The $\mu$TPC analysis, proven to improve the spatial resolution in the case of $^{10}$B converters, is extended to gadolinium based detectors. For the first time, a Gd-GEM was successfully operated to detect neutrons with a measured efficiency of 11.8\% at a wavelength of 2~\AA~and a position resolution better than 250~$\mu$m.}


\begin{document}
\section{Introduction}\label{sec:intro}
The European Spallation Source (ESS)~\cite{ESS} in Lund/Sweden is foreseen to start operations in 2019 and will become the world's most powerful thermal neutron source. Its brightness will be significantly higher than the brightness at existing reactor sources like the Institut Laue-Langevin (ILL)~\cite{ILL} or other spallation sources like the Spallation Neutron Source (SNS)~\cite{SNS} and the Japan Proton Accelerator Research Complex (J-PARC)~\cite{J-PARC}. 22 neutron scattering instruments are currently planned as the baseline suite for the facility~\cite{ESS_TechnicalDesignReport}. For each of the instruments efficient thermal neutron detectors are a crucial component~\cite{Vertex}.
Traditionally the dominant detector technology at neutron scattering instruments has been gaseous $^3$He detectors~\cite{ILL_Neutron}.
In light of the $^3$He crisis~\cite{He3_crisis1, He3_crisis2} an extensive international R\&D program is currently under way in order to develop efficient and cost-effective detectors based on other isotopes~\cite{ICND,Zeitelhack, Guerard}. Solutions like gas proportional counters surrounded by thin films of $^{10}$B-enriched boron carbide~\cite{B4cfilms,B10_multigrid, Bigault} have been developed for instruments with large area detectors. 

The ESS Macromolecular Diffractometer (NMX)~\cite{ESS_TechnicalDesignReport} requires three 60 x 60~cm$^{2}$ detectors with reasonable detection efficiency and ca 200 $\mu$m spatial resolution. At reactor sources macromolecular crystallography instruments typically use neutron image plates~\cite{ImagePlate} with this resolution, but spallation source instruments require time resolution that the image plates lack altogether. Scintillation based detectors~\cite{IBIX, MANDI} are currently limited to ca 1 mm spatial resolution. 

For spallation source instruments solid neutron converters in combination with Micro Pattern Gaseous Detectors (MPGDs)~\cite{MPGD} are a promising option to achieve the spatial resolution of the neutron image plate with time resolution, high-rate capabilities and a good neutron detection efficiency. 
A data analysis technique based on the principle of the Time Projection Chamber (TPC) was developed, and has already been successfully applied to MPGDs with $^{10}$B$_4$C converters. A position resolution better than 200~$\mu$m was obtained by determining the start of the ionization track~\cite{uTPC_Boron}, but the detection efficiency was lower than 5$\%$. This so-called $\mu$TPC analysis~\cite{uTPC} is now extended to the case of Gas Electron Multipliers (GEM) detectors with more efficient gadolinium converters. This work presents the measurements carried out at the R2D2 beam line at the JEEP II research reactor at the Institute for Energy Technology (IFE) in Kjeller/Norway~\cite{IFE}. It consists of a description of the design, the setup, the measurement and the data analysis that was performed in order to retrieve the full information on the position of the neutron conversion.

\section{Gadolinium as neutron converter}
\label{sec:neutron converter}

\subsection{Conversion electrons}
\label{sec:Conversion electrons}

\begin{figure}[htbp]
\centering
\subfloat[In Gd before interaction with converter\label{fig: spectrum_ekin_converter}]{
\includegraphics[width=.49\textwidth]{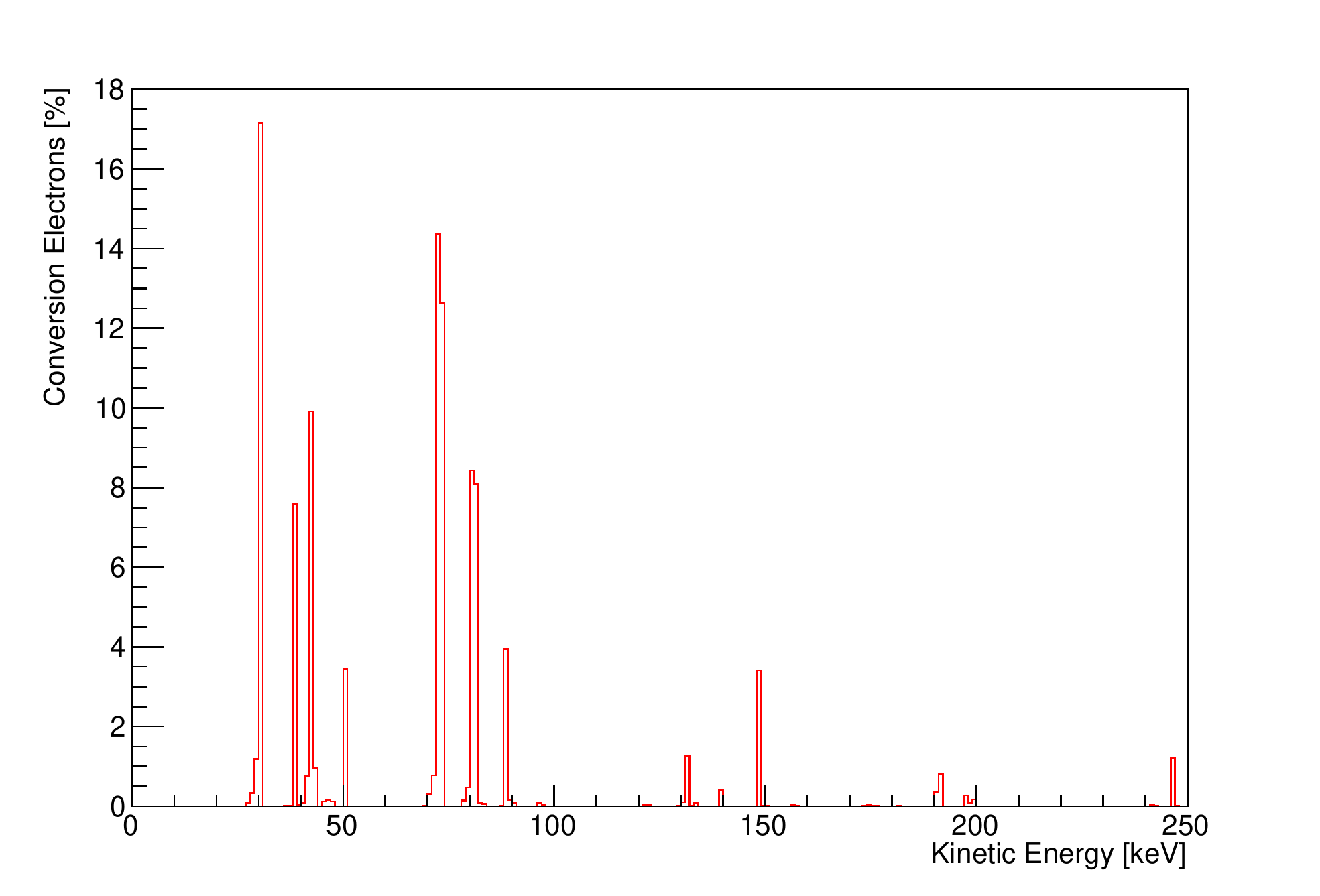}%
}
\subfloat[Leaving converter (backwards configuration)\label{fig: spectrum_ekin_drift}]{
\includegraphics[width=.49\textwidth]{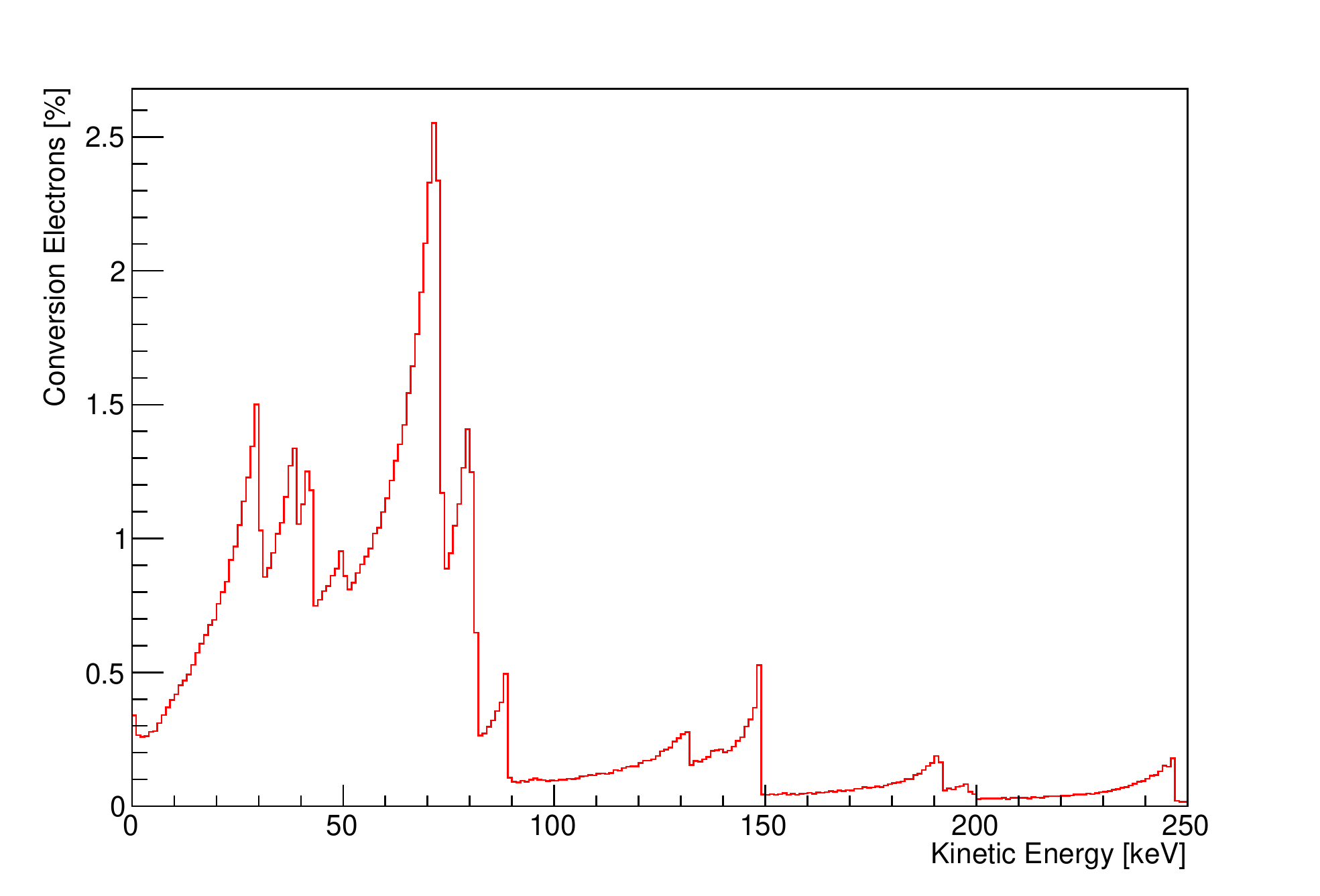}%
}
\caption{Kinetic energy of conversion electrons (bin size 1 keV) as simulated with Geant4.}
\label{fig: spectra_ekin}
\end{figure}

Despite the very large neutron capture cross section of $^{155}$Gd and $^{157}$Gd, the material is not a popular converter due to the nature and the energy of the secondary particles. In fact, after the neutron capture, gadolinium releases prompt $\gamma$s with an energy of up to 9 MeV and conversion electrons with energies ranging from 29 keV to 250 keV. Figure~\ref{fig: spectra_ekin} shows the simulated kinetic energy distribution of the conversion electrons after the neutron capture in a 250~$\mu$m thick natural Gd converter, and at the moment when they are leaving the converter. Natural gadolinium contains 14.80$\%$ of $^{155}$Gd and 15.65$\%$ of $^{157}$Gd, the remainder are Gd isotopes without significant cross section for thermal neutrons. Since the conversion electrons loose energy in the Gd, the discrete energy distribution present in the converter smears out and turns into a more continuous spectrum when the conversion electrons leave the converter. The mean kinetic energy of the conversion electrons escaping from the converter is 70~keV. 

Table \ref{table: ranges} shows the Continuous Slowing Down Approximation (CSDA) range and the maximum penetration depth of electrons between 10~keV and 250~keV in Ar/CO$_2$ 70/30 at Normal Temperature and Pressure (NTP)\footnote[1]{Normal Temperature and Pressure is defined as a temperature of 293.15 K and a pressure of 1 atm.}. Whereas the CSDA range is defined as the mean path length of a charged particle in an absorber, the maximum penetration depth is defined as the depth in the absorbing medium beyond which no particles are observed to penetrate~\cite{MPD}. In the gaseous detector under study, the absorbing material is the 10~mm long drift volume filled with Ar/CO$_2$ 70/30~\ref{fig: detector}. The conversion electron track is fully contained in the drift space if its maximum penetration depth is smaller than the drift length. Conversion electrons between 10~keV and 250~keV have a maximum penetration depth between 1~mm and 26~cm. Therefore, depending on the initial position and angle and the size of the drift volume, only a fraction of the conversion electrons tracks will be fully contained in the detector.

\begin{table}
\centering
\footnotesize
\begin{threeparttable}
\begin{tabular}[htbp]{ r|r|r|r|}
\cline{1-4}
\multicolumn{1}{ |r| }{Kinetic Energy [keV]} & \multicolumn{1}{ |r| }{CSDA Range [cm]\tnote{a}} & \multicolumn{1}{ |r| }{Maximum Penetration Depth [cm]\tnote{b}} & \multicolumn{1}{ |r| }{Standard Deviation [cm]\tnote{c}} \\ 
\hline
\multicolumn{1}{ |r|}{10} & 0.20 & 0.09 & 0.03 \\ 
\multicolumn{1}{ |r|}{20} & 0.66 & 0.33 & 0.09 \\ 
\multicolumn{1}{ |r|}{30} & 1.34 & 0.69 & 0.19 \\ 
\multicolumn{1}{ |r|}{40} & 2.21 & 1.16 & 0.32 \\ 
\multicolumn{1}{ |r|}{50} & 3.26 & 1.73 & 0.47 \\ 
\multicolumn{1}{ |r|}{60} & 4.46 & 2.39 & 0.64 \\ 
\multicolumn{1}{ |r|}{70} & 5.81 & 3.13 & 0.83 \\ 
\multicolumn{1}{ |r|}{80} & 7.29 & 3.95 & 1.04 \\ 
\multicolumn{1}{ |r|}{90} & 8.90 & 4.85 & 1.28 \\ 
\multicolumn{1}{ |r|}{100} & 10.62 & 5.80 & 1.53 \\ 
\multicolumn{1}{ |r|}{125} & 15.40 & 8.51 & 2.21 \\ 
\multicolumn{1}{ |r|}{150} & 20.76 & 11.49 & 2.94 \\ 
\multicolumn{1}{ |r|}{175} & 26.61 & 14.74 & 3.79 \\ 
\multicolumn{1}{ |r|}{200} & 32.90 & 18.33 & 4.62 \\ 
\multicolumn{1}{ |r|}{250} & 46.55 & 26.18 & 6.57 \\ 
\hline
\end{tabular}
\caption{Path length and maximum penetration depth of electrons in Ar/CO$_2$ 70/30 at NTP.}
\label{table: ranges}
\begin{tablenotes}
\item[a] Continuous slowing down approximation (CSDA) range from ESTAR database~\cite{ESTAR}.
\item[b,c] Maximum Penetration Depth simulated with Geant4 Photoabsorption Ionization Model.
\end{tablenotes}
\end{threeparttable}
\end{table}

\begin{figure}[htbp]
\centering
\subfloat[Simulated spectrum of conversion electrons\label{fig: spectrum_drift}]{
\includegraphics[width=.49\textwidth]{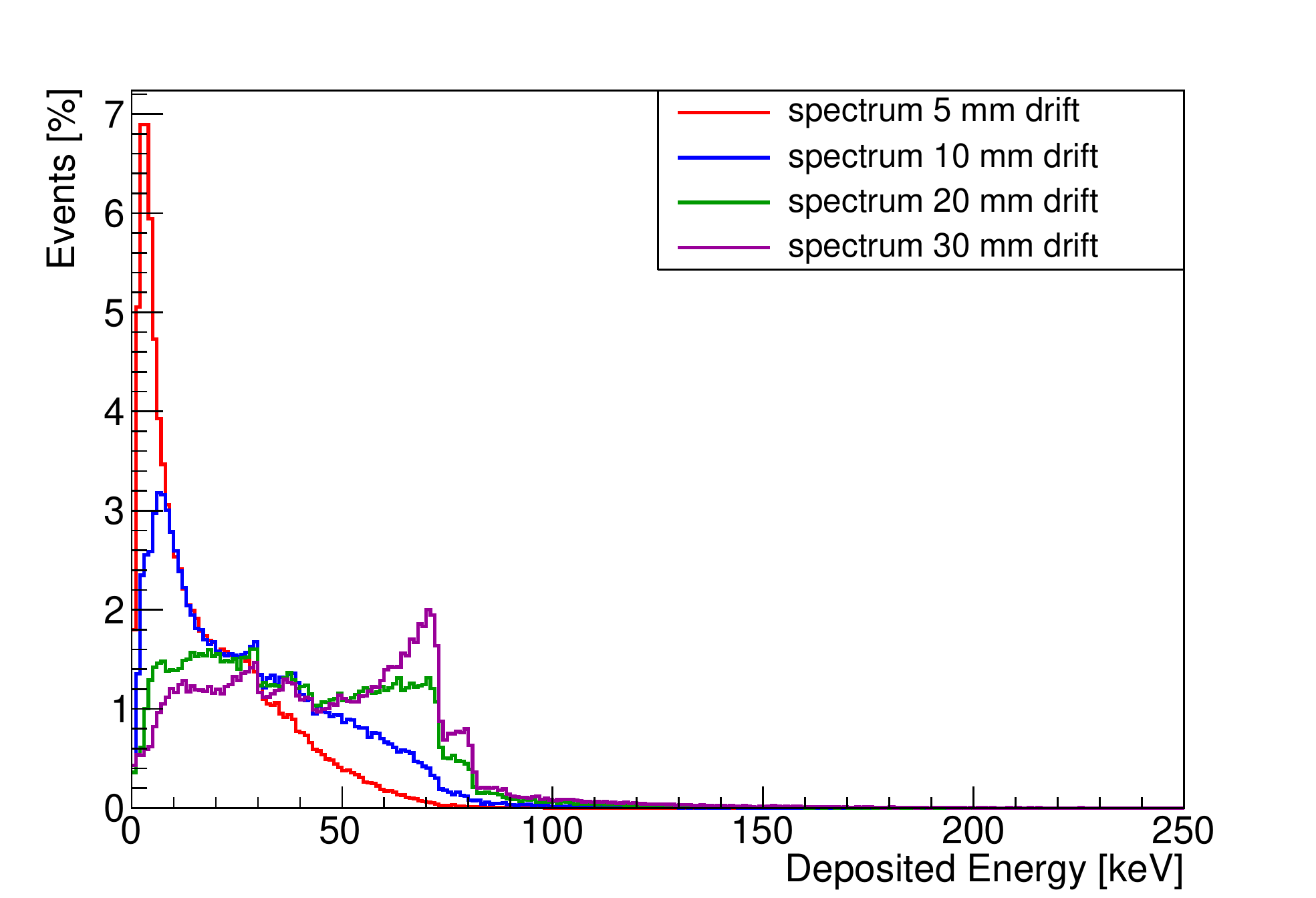}%
}
\subfloat[Deposited energy in drift\label{fig: deposited_energy}]{
\includegraphics[width=.49\textwidth]{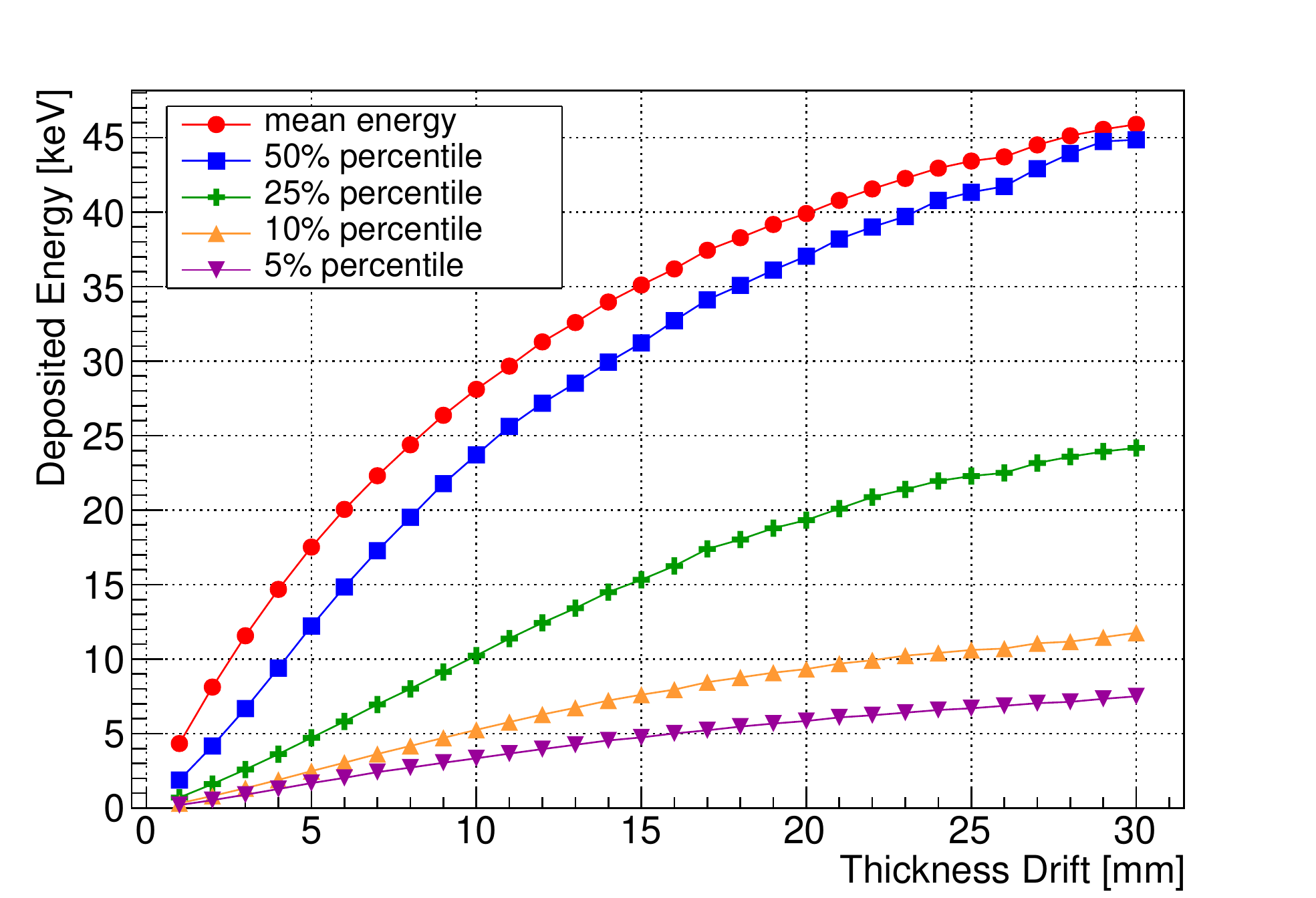}%
}
\caption{Simulated spectrum of conversion electrons in drift (bin size 1 keV) and deposited energy in drift depending on the drift space .}
\label{fig: drift}
\end{figure}

Figure \ref{fig: drift} confirms that the energy deposit of the conversion electrons highly depends on the size of the drift. To detect the 70~keV peak that dominates the spectrum in figure \ref{fig: spectrum_ekin_drift} in the spectrum of the deposited energies, the drift space would have to be around 30~mm long. For a 10~mm drift, only tracks of conversion electrons with an energy of up to 30~keV will be fully contained. Figure \ref{fig: deposited_energy} depicts the mean deposited energy and the 5$\%$, 10$\%$, 25$\%$ and 50$\%$ percentiles depending on the drift size. For a 10~mm drift, Gd conversion electrons on average deposit only 28 keV in the drift space by ionizing the Ar/CO$_2$. The most probable energy deposition is 24~keV, and an energy threshold of 5~keV will cut away 10$\%$ of the conversion electron signal.

The deposited energy of the conversion electrons is thus small when compared to the secondaries from $^3$He, $^6$Li, and $^{10}$B converters. In addition to the small amount of deposited energy, gadolinium is also a high Z material. Gadolinium based detectors are therefore more sensitive to gamma background, when only the signal amplitude is used for the signal discrimination. The gamma sensitivity and advanced discrimination techniques though will not be discussed in this paper.  The ranges of the conversion electrons also seem to be in contrast with the spatial resolution requirements. In the past, a CsI layer applied to the gadolinium was used as an electron emitter to localize the electron energy loss and, therefore, improve the spatial resolution~\cite{HZB1, HZB2}. The disadvantage in this case is the very small amount of primary electrons and deposited energy that requires a detector with very large gain. 

\begin{figure}[htbp]
\centering
\subfloat[3D\label{fig: electron_3D}]{
\includegraphics[width=.34\textwidth]{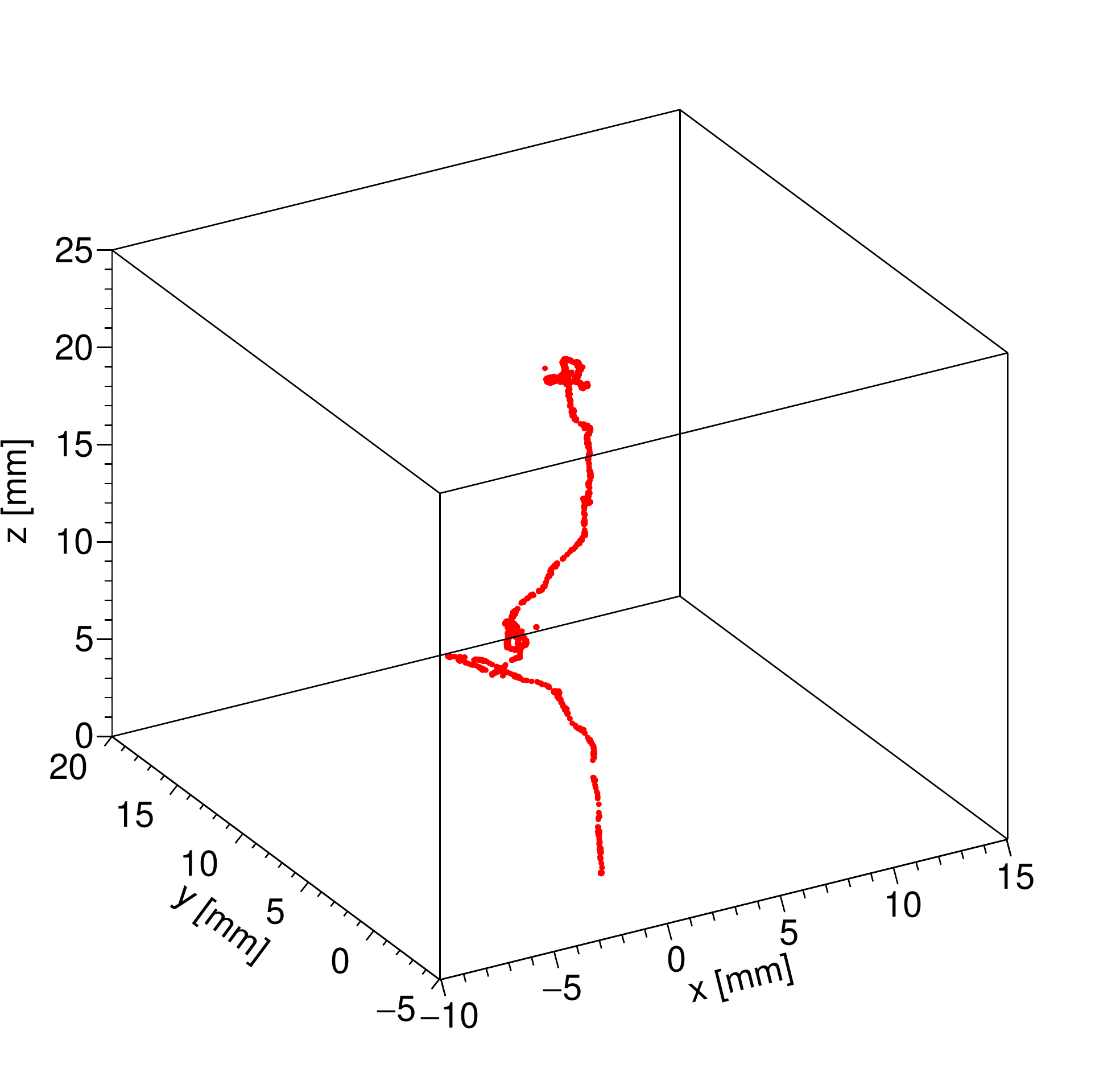}%
}
\subfloat[Projection zx\label{fig: electron_zx}]{
\includegraphics[width=.32\textwidth]{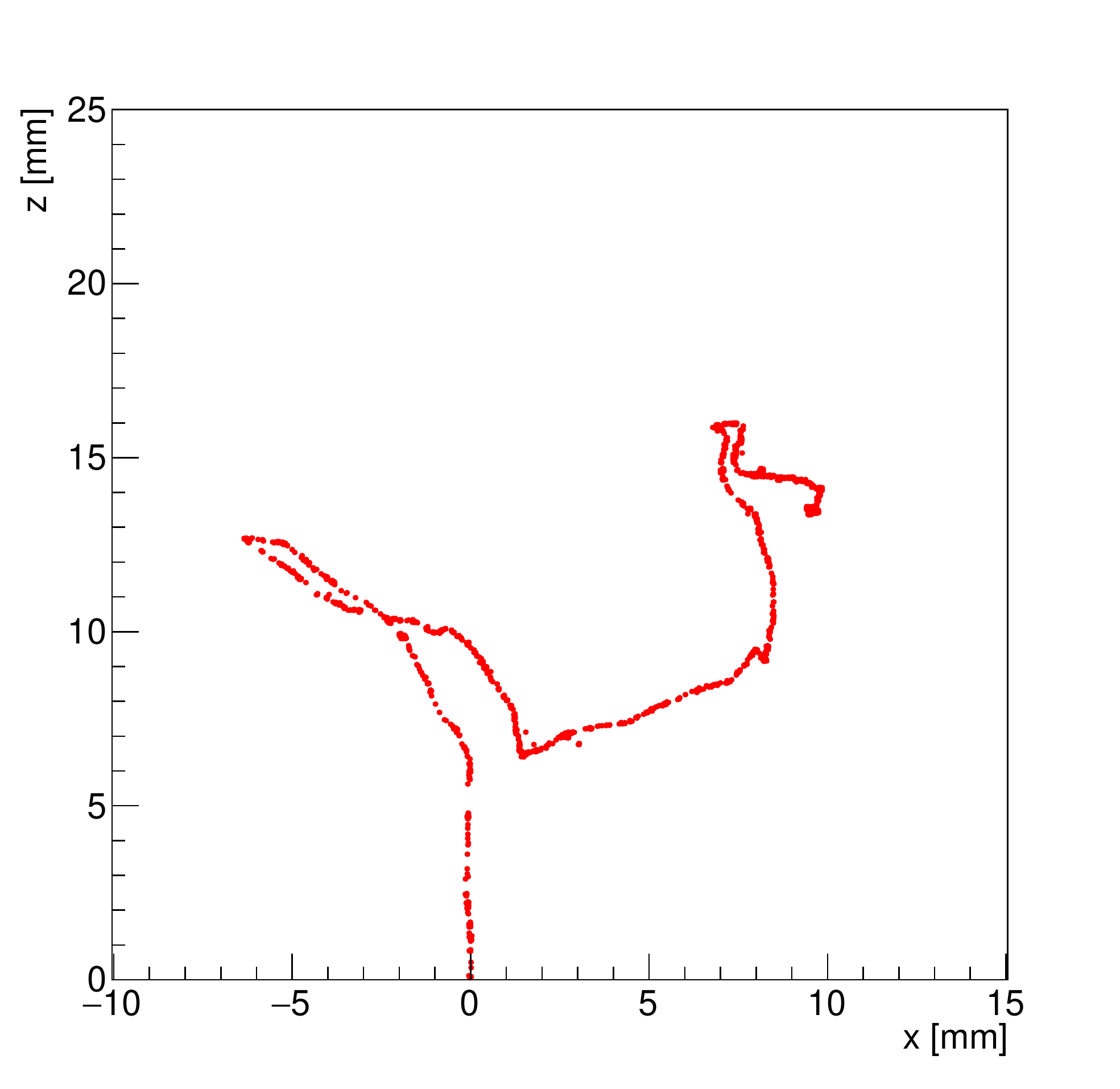}%
}
\subfloat[Projection zy\label{fig: electron_zy}]{
\includegraphics[width=.32\textwidth]{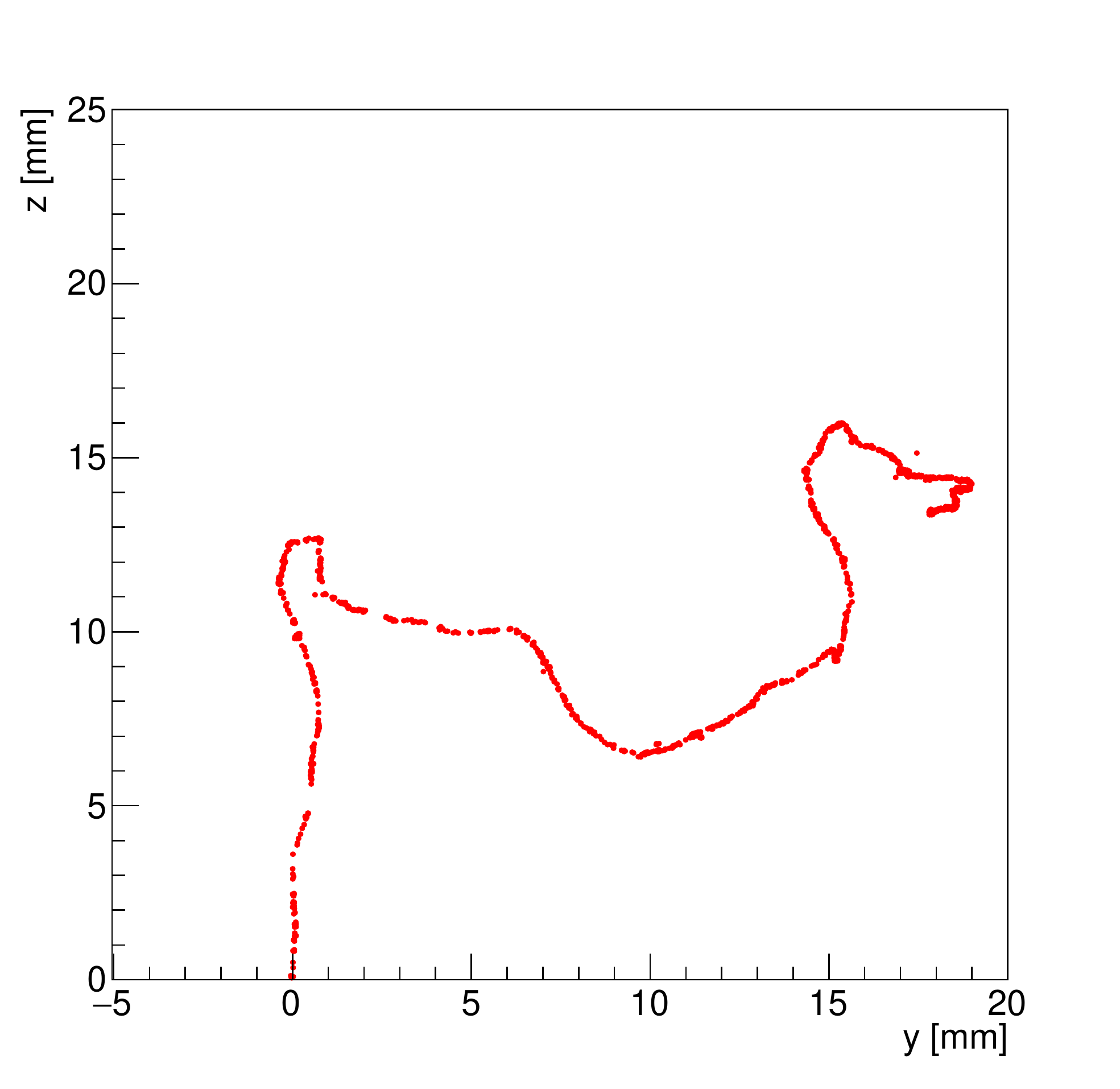}%
}
\caption{Example track of 70~keV electron in Ar/CO$_2$ 70/30 originating in (0,0,0) as simulated with Degrad.}
\label{fig: electron}
\end{figure}

In this work a different approach was followed: the conversion electrons from the neutron capture ionize a macroscopic portion of the active volume, and an offline analysis takes care of the reconstruction of the neutron interaction point from the three dimensional topology of the event. The challenge is here that in contrast to the $\alpha$ particles and Li ions created in the $^{10}$B$_4$C converter, the gadolinium conversion electrons have not only on average a far larger range, but also do not leave a straight ionization track. Figure~\ref{fig: electron} shows the typical curved track of a 70~keV electron starting at (0,0,0) in Ar/CO$_2$ 70/30 as simulated with Degrad~\cite{Degrad}. The electron energy will not be fully deposited in a detector with a 10~mm drift space, as already illustrated in table~\ref{table: ranges}. Further, it is apparent that the path length of the conversion electrons is longer than their maximum penetration depth due to multiple scattering.

\subsection{Converter efficiency}
\label{sec:Converter efficiency}

The Gd and detector simulations were carried out with Geant4~\cite{Geant4a} version 10.1 using the high-precision neutron data G4NDL4.5.PUB, the gamma level data PhotonEvaporation3.1, the flag G4NEUTRON\-HP{\_}\-USE{\_}ONLY\-{\_}PHOTONEVAPORATION and NTP conditions. The wavelength of the neutrons in the simulation was set to 2~\AA, and the used geometry is depicted in figure~\ref{fig: detector}. Given the large thickness of the cathode, the converter was used in \emph{backwards configuration}, i.e. the neutron beam, impinging orthogonally to the detector, crossed the readout board and the GEMs before reaching the gadolinium cathode. In this way the conversion electrons do not need to traverse the entire gadolinium thickness in order to reach the active volume, which leads to a higher neutron detection efficiency. On the other hand the scattering of the neutrons in the material of the readout decreases the position resolution and the efficiency. Geant4 does only contain thermal scattering data for about 20 different atoms and materials. One of the materials for which scattering data based on the evaluated nuclear data files ENDF/B-VII exists, is the hydrogen atom in Polyethylene. This atom is called TS{\_}H{\_}of{\_}Polyethylene in the Geant4 nomenclature. Therefore, to consider the scattering of the cold neutrons, the hydrogen atoms in the Kapton of the GEM foils and the FR4 of the readout board were defined as being of type TS{\_}H{\_}of{\_}Polyethylene.  If further the G4NeutronHP\-Thermal\-Scattering model is registered as model for the neutron elastic physics process, Geant4 will use the scattering data in the simulation. 

The simulations of the detector under test indicate that due to the scattering only 81.3$\%$ of the 2~\AA~neutrons reach the converter, where all of them are captured. The number of conversion electrons per captured neutron amounts to 0.656, which agrees quite well with the existing literature. A number of 59.9 conversion electrons per 100 captured neutrons in natural Gd can be found in the work of Harms and McCormack~\cite{Harms}, whereas Hargrove et al. based on the gamma level schemes in the work of B\"{a}cklin et al.~\cite{Backlin} and Greenwood et al.~\cite{Greenwood} give the theoretical number of conversion electrons per captured neutron as 0.66~\cite{Hargrove}. To calculate the neutron detection efficiency of the detector, not the number of conversion electrons per captured neutron is relevant, but the probability that at least one conversion electron is produced and reaches the drift. The probability for the production of at least one conversion electron in the converter amounts to 51.1$\%$. This number is lower than the number of conversion electrons per 100 neutrons, because in some cases more than one conversion electron is produced during the neutron capture. In backwards mode, the probability of at least one conversion electron reaching the drift space is 14.2$\%$. Assuming that each conversion electron that reaches the drift can be detected, the detection efficiency of the detector is 14.2$\%$ in backwards configuration. Taking into consideration the scattering, the detection efficiency is 0.813 times 0.142 or 11.5$\%$. 

Bruckner et al. calculated the detection efficiency for a Si detector with a natural Gd converter of 25~$\mu$m thickness in backwards configuration~\cite{Bruckner}. For neutrons with 1.8~\AA~wavelength, the calculated efficiency was around 11.6$\%$. According to Abdushukurov, the calculated detection efficiency for a 100~$\mu$m thick natural Gd converter in backwards mode with neutrons of 1.8~\AA~wavelength amounts to 15.2$\%$~\cite{Abdushukurov}. With 1.8~\AA~neutrons and without scattering of the neutrons, Geant4 arrives in backwards configuration at an detection efficiency of 13.9$\%$ for a converter thickness of 25~$\mu$m and 14.1$\%$ for a converter thickness of 100~$\mu$m. The simulated efficiencies are thus slightly lower than the efficiency calculated by Abdushukurov and slightly higher than the efficiency calculated by Bruckner et al., but agree reasonably well with the literature. In general, there seems to be the tendency that measured efficiencies are higher than calculated efficiencies. The measured efficiency by Bruckner et al. for various wavelengths was considerably higher than their calculation, e.g. for 1.7~\AA~neutrons 19$\%$ were measured with a 25~$\mu$m thick converter. The authors explain the discrepancy with the partial detection of the X-rays and gamma rays that are produced during the neutron capture. Hargrove et al. derive from their measurements with thermal neutrons an impressive total efficiency (backwards and forwards mode) of 63$\%$ for a 4~$\mu$m thick foil of natural Gd. They try to explain the discrepancy with contributions from higher gamma levels that are usually not included in the theoretical calculations of the conversion electron spectrum.

\section{Experimental setup}
\label{sec:setup}
\subsection{R2D2 beam line}
\label{subsec:R2D2 beam line}
\begin{figure}[htbp]
\centering
\subfloat[Detector under test at R2D2 beam line\label{fig: beamline}]{
\includegraphics[width=.45\textwidth]{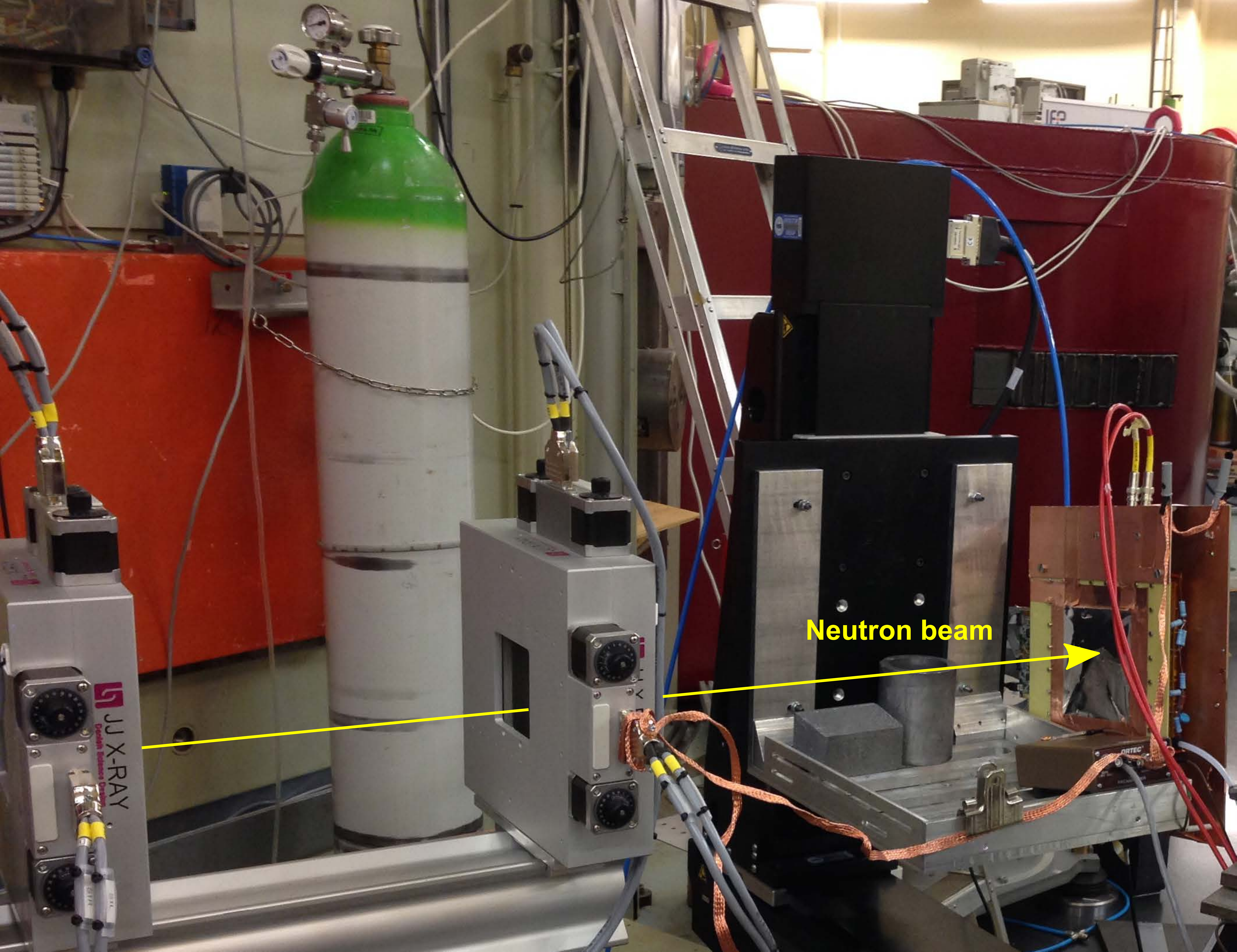}%
}
\subfloat[Schematic representation of the detector\label{fig: detector}]{
\includegraphics[width=.53\textwidth]{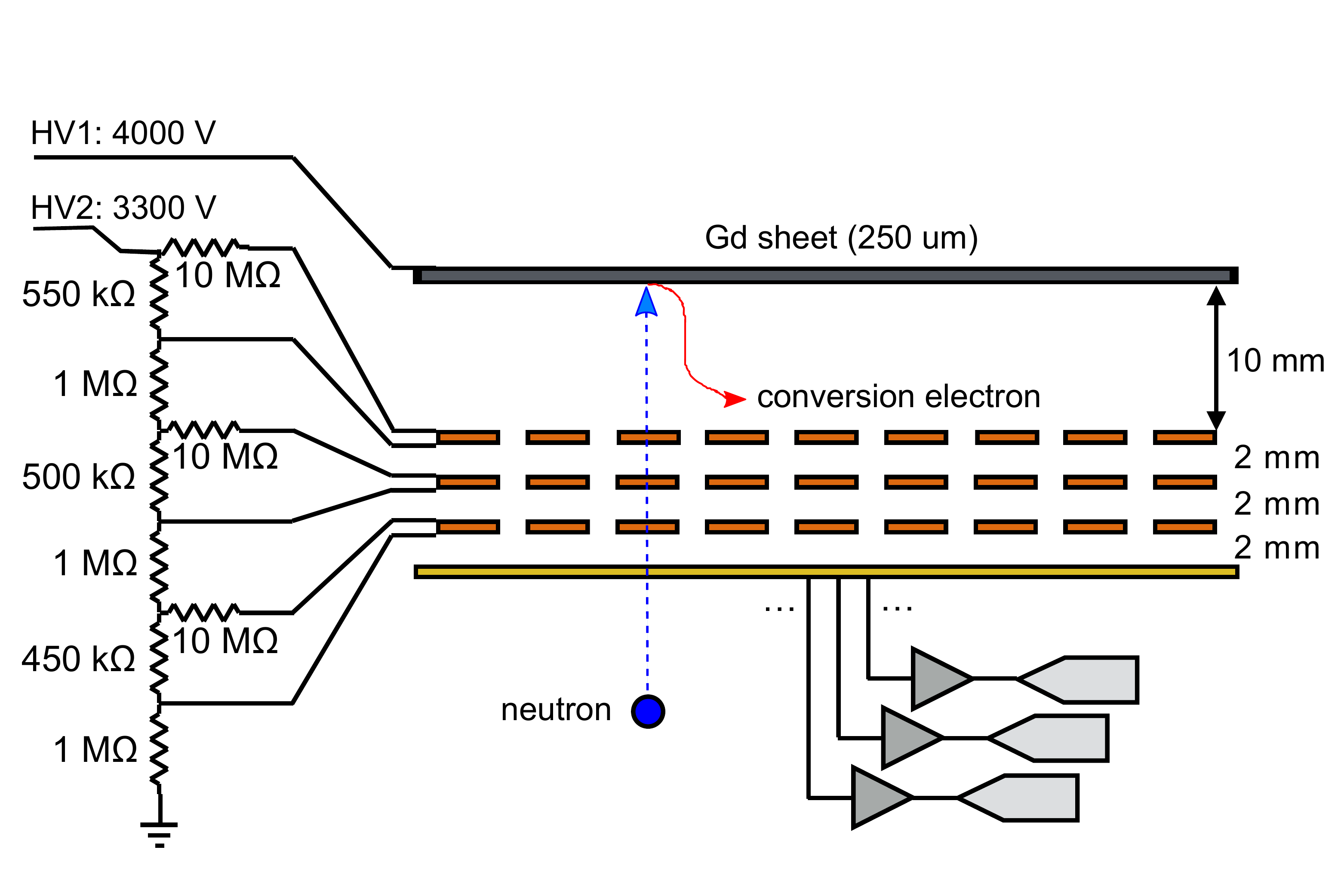}%
}
\caption{Measurement setup at R2D2 beam line with collimation slits and Triple-GEM detector in backwards configuration.}
\label{fig: setup}
\end{figure}

For the detector tests at JEEP II, the monochromator of the R2D2 beam line was optimized to deliver mono-energetic neutrons of 2~\AA. A multi-slit Gd$_{2}$O$_{3}$ Soller collimator that limits the angular divergence of the beam to 30' is inserted into the R2D2 shutter. Figure \ref{fig: beamline} shows the beam line behind the shutter. It was equipped with a beam monitor, two sets of movable collimating slits made of borated Al blades, and a calibrated Reuter-Stokes 1 inch $^3$He tube (model RS-P4-0810-220). The tube was operated at a $^3$He pressure of 10 atm, and a voltage of 1400~V. The walls of the tube are made of stainless steel and have a thickness of 0.508~mm. At a wavelength of 2.5~\AA, the efficiency of the tube is (96.3 $\pm$ 0.3)$\%$. From this efficiency, the efficiency for neutrons with a wavelength of 2~\AA~was extrapolated to (95 $\pm$ 0.5) $\%$.

\subsection{Detector setup, readout and DAQ}
\label{subsec:Detector setup, readout and DAQ}
The detector setup is shown in figure~\ref{fig: detector}. The detector under test was a \textit{standard} 10 x 10~cm$^{2}$ Triple-GEM~\cite{Triple-GEM} detector flushed with Ar/CO$_2$ 70/30 mixture at 10~l/h at room temperature and atmospheric pressure. It consisted of three GEM foils~\cite{GEM} stacked at a distance of 2~mm from each other and powered with a resistor chain. The detector was operated at an effective gain of about 5000. The conversion volume or drift space was 10~mm long, and as cathode served a 250 $\mu$m thick sheet of natural gadolinium. The high voltage to the cathode was provided with a second power supply. A drift field of 700 V/cm was chosen to avoid electron attachment to electronegative impurities and the subsequent loss of primary ionization electrons, while keeping the drift velocity smaller than 2.0~cm/$\mu$s~\cite{Drift} at atmospheric pressure. The readout board was a Cartesian x/y strip readout~\cite{Triple-GEM, GEM_readout1} with 256 strips and 400~$\mu$m pitch in x and y direction. Four APV-25~\cite{APV} hybrid chips per readout were used to pre-amplify the signals. The APVs were operated in multi mode, and the signal from the bottom of the third GEM foil served as external trigger signal. In APV multi mode, three consecutive pipeline columns are reserved each time the chip is triggered. Each column is then read out separately in peak readout mode~\cite{APV_mode}. The waveforms were digitized with the Scalable Readout System (SRS)~\cite{SRS} and acquired with the ALICE DAQ system DATE~\cite{DATE} and the Automatic MOnitoRing Environment (AMORE)~\cite{AMORE} software. 

\subsection{Detection efficiency}
\label{Detection efficiency}

The efficiency measurement of the detector was carried out with a 0.5 x 0.5~cm$^{2}$ collimated beam. The beam was collimated using a 2~cm thick $^{10}$B$_4$C collimator with an opening of 0.5 x 0.5~cm$^{2}$ that was placed at a distance of 5~cm in front of the detector. First the $^3$He tube was installed directly behind the collimator in the 5~cm gap in front of the detector as shown in figure~\ref{fig: setup_efficiency}. With the beam switched on, a rate of (31.4 $\pm$ 0.02)~kHz was measured with the $^3$He tube. With either a non-paralyzable or a paralyzable model for the dead time behaviour of the $^3$He tube electronics, a 3.2~$\%$ correction of the measured rate is required. Consequently the corrected neutron flux amounted to (32.4 $\pm$ 0.02)~kHz. After the removal of the $^3$He tube, the detector measured a rate of (10.88 $\pm$ 0.10)~kHz. Without beam, the detector measured a negligible background rate of (120.0 $\pm$ 11)~Hz uniformly distributed over the full detector surface. For the efficiency measurements the trigger threshold was set to 2~keV, to be far away from the noise level and to avoid any over estimation of the neutron efficiency. 

To determine which parts of the measured detector rate were caused by beam induced background and which by the neutron signal, the data obtained with the x/y strip readout of the detector was analysed. A 2D Gaussian has been chosen to represent the beam induced background. For the neutron signal, a Gaussian form has been assumed for the point spread function, resulting in the intensity distribution function
\[ I(x,y) = \frac{I_{0}}{16} \;  \textrm{erfc}\left(\frac{-x+x_{1}}{\sqrt{2} \; \sigma_{1}}\right) \; \textrm{erfc}\left(\frac{x-x_{2}}{\sqrt{2} \; \sigma_{2}}\right) \; \textrm{erfc}\left(\frac{-y+y_{1}}{\sqrt{2} \; \sigma_{3}}\right) \; \textrm{erfc}\left(\frac{y-y_{2}}{\sqrt{2} \; \sigma_{4}}\right).\]
Here x and y are the reconstructed $x$/$y$ positions of the hits (see chapter ~\ref{sec:analysis}). $I_{0}$ is the neutron intensity at the center of the beam spot. The parameters $x_{1}$ and $y_{1}$ are the positions of the rising edges, whereas $x_{2}$ and $y_{2}$ are the positions of the falling edges. $\sigma_{1}$ to $\sigma_{4}$ are the width parameters of the Gaussian point spread functions associated with each edge location. Minuit2~\cite{Minuit2} was used to fit the sum of the 2D Gaussian and the intensity function to the data with the reconstructed $x$/$y$ positions. From the integral of the neutron intensity distribution, it was thus determined that (4.01 $\pm$ 0.02)~kHz of the 10.88~kHz were due to cold neutrons. This leads to a measured neutron flux of about 8~kHz/cm$^{2}$ in an area of about 0.5~cm$^{2}$, a background flux of about 640~Hz/cm$^{2}$ in an area of about 10.8~cm$^{2}$ and thus a signal to background ratio of about 12 to 1.

Using these values, the measured efficiency of the detector amounted to 
\[(4.01 \pm 0.02)~kHz \times (95 \pm 0.5) \% / (32.4 \pm 0.02)~kHz = (11.8 \pm 0.1)\%.\]

To compare the measured efficiency with the simulation, the 2~keV threshold used during the measurements also has to be applied to the simulated efficiency. When applying this correction, the simulated efficiency drops from  11.5$\%$ to 11$\%$. The discrepancy of 0.8$\%$ could be due to an overestimation of the scattering in the simulation.

\subsection{Position resolution}
\label{subsec:Position resolution}

\begin{figure}[htbp]
\centering
\subfloat[Setup for efficiency measurements\label{fig: setup_efficiency}]{
\includegraphics[width=.45\textwidth]{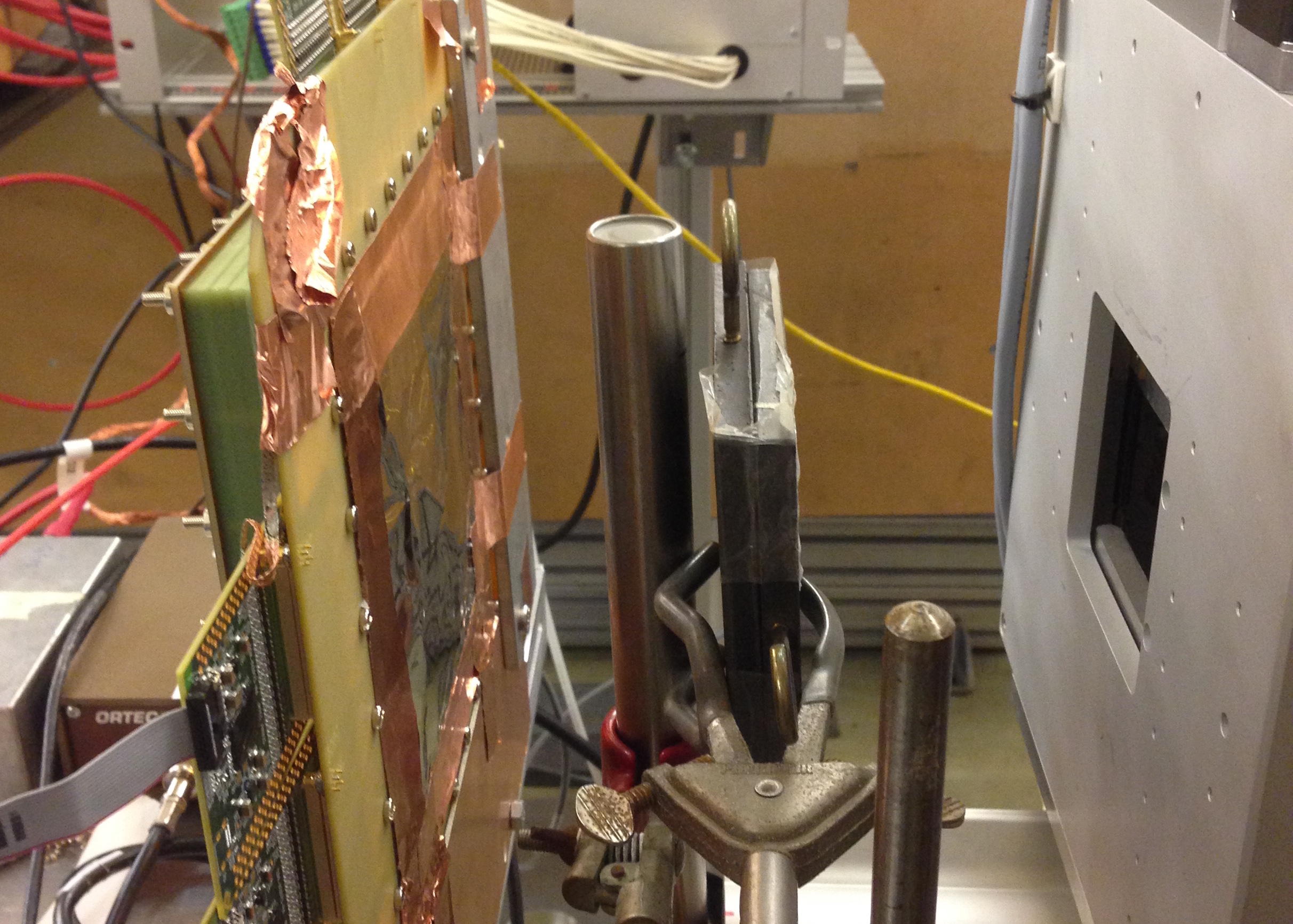}%
}
\subfloat[Simulated and measured spectrum of all particles\label{fig: spectrum_measured}]{
\includegraphics[width=.53\textwidth]{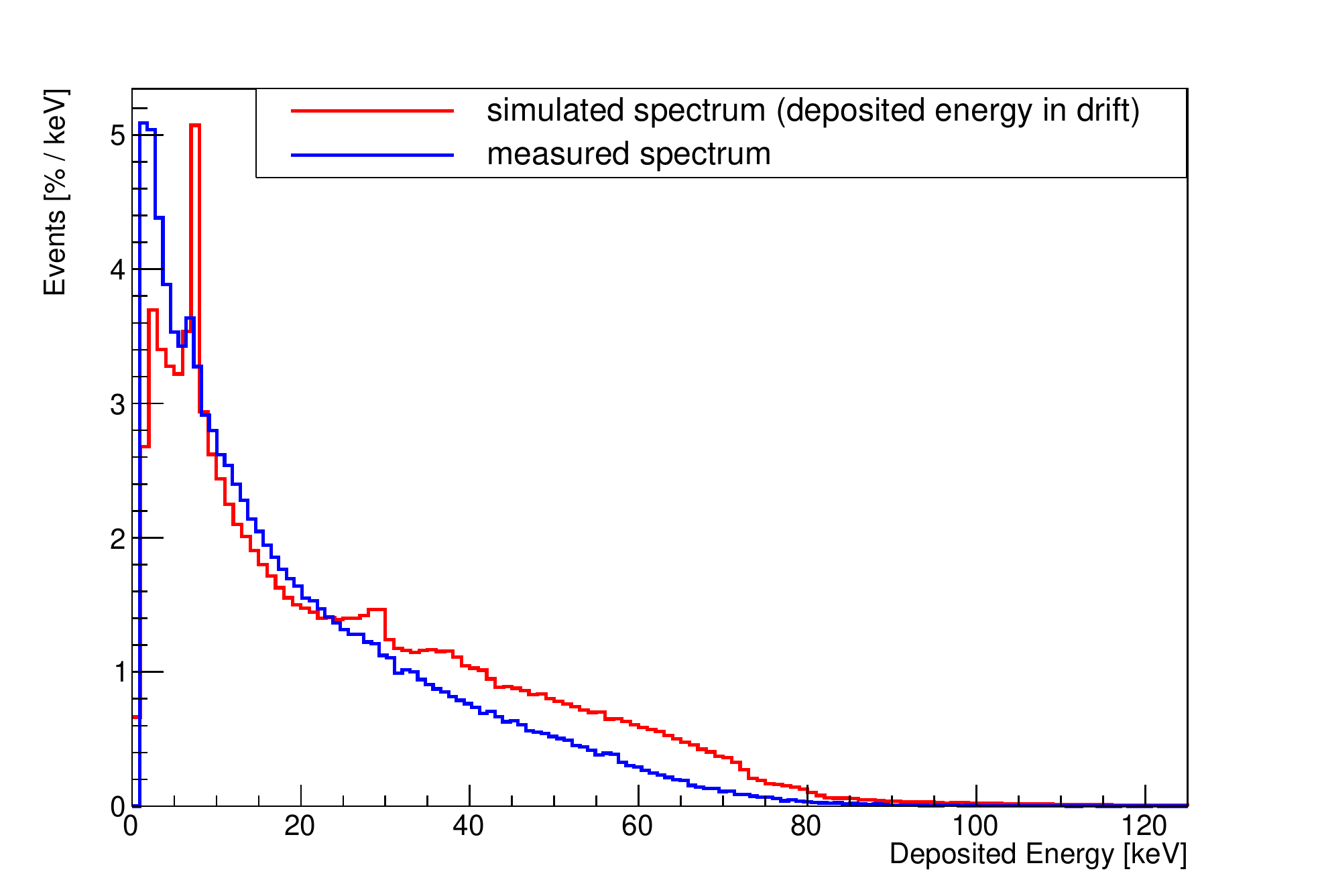}%
}
\caption{Setup for the measurement of the detector efficiency and comparison of simulated and measured spectrum in drift (bin size 1 keV).}
\label{fig: spectra}
\end{figure}

For the analysis of the position resolution, a data set of 90000 events was acquired with a collimated beam of 0.3 x 10~cm$^{2}$. As shown in figure \ref{fig: hit_distribution_3mm_10cm}, the detector clearly recorded the shape of the beam. During this measurement, the detector happened to be rotated by 7$^{\circ}$. About 10$\%$ of the events in the data set were excluded from the analysis due to electrical problems like misfiring strips. A comparison of the acquired pulse height spectrum with the simulated spectrum of the deposited energy in the drift volume (figure~\ref{fig: spectrum_measured}) shows a similar shape. The simulation was carried out using the Geant4/Garfield++ interface~\cite{Geant4a,Garfield++,Interface}. With the help of the interface,  a complete neutron detector simulation is possible. But to simplify things and speed up the simulation, the amplification stages and the charge collection in the read out have not been simulated here. Therefore the simulated spectrum shows far better energy resolution than a real spectrum. To calibrate the neutron pulse height spectrum, first a pulse height spectrum with a $^{241}$Am source was acquired. Whereas the 59.9~keV americium gamma was not directly visible, the spectrum clearly showed the 8~keV copper X-ray peak which stems from the interaction of the Am gammas with the copper of the GEM foils. The 8~keV copper X-ray peak was then used for the energy calibration of the neutron spectrum. In the neutron spectrum, the conversion electron peaks at 29~keV and 39~keV are too small to be detected within the energy resolution of the detector. The 8~keV copper X-ray peak on the other hand is very small but also visible in the neutron spectrum. It is caused by prompt gammas from the Gd neutron capture that again interact with the copper of the GEM foils. During the measurements, the discriminator threshold for the trigger signal from the bottom of the third GEM foil was set to 1~keV. As already shown in table \ref{table: ranges} and figure \ref{fig: spectrum_drift}, the maximum penetration depth of the conversion electrons with energies higher than 100~keV is so large, that their tracks are never fully contained in the detector volume. Therefore they deposit less energy in the drift space than the conversion electrons with lower energies. The spectrum thus does basically not show any energy deposit larger than 100~keV. 

\section{$\mu$TPC analysis}
\label{sec:analysis}
\begin{figure}[htbp]
\centering
\subfloat[strip 1\label{fig: time_structure_1}]{
\includegraphics[width=.32\textwidth]{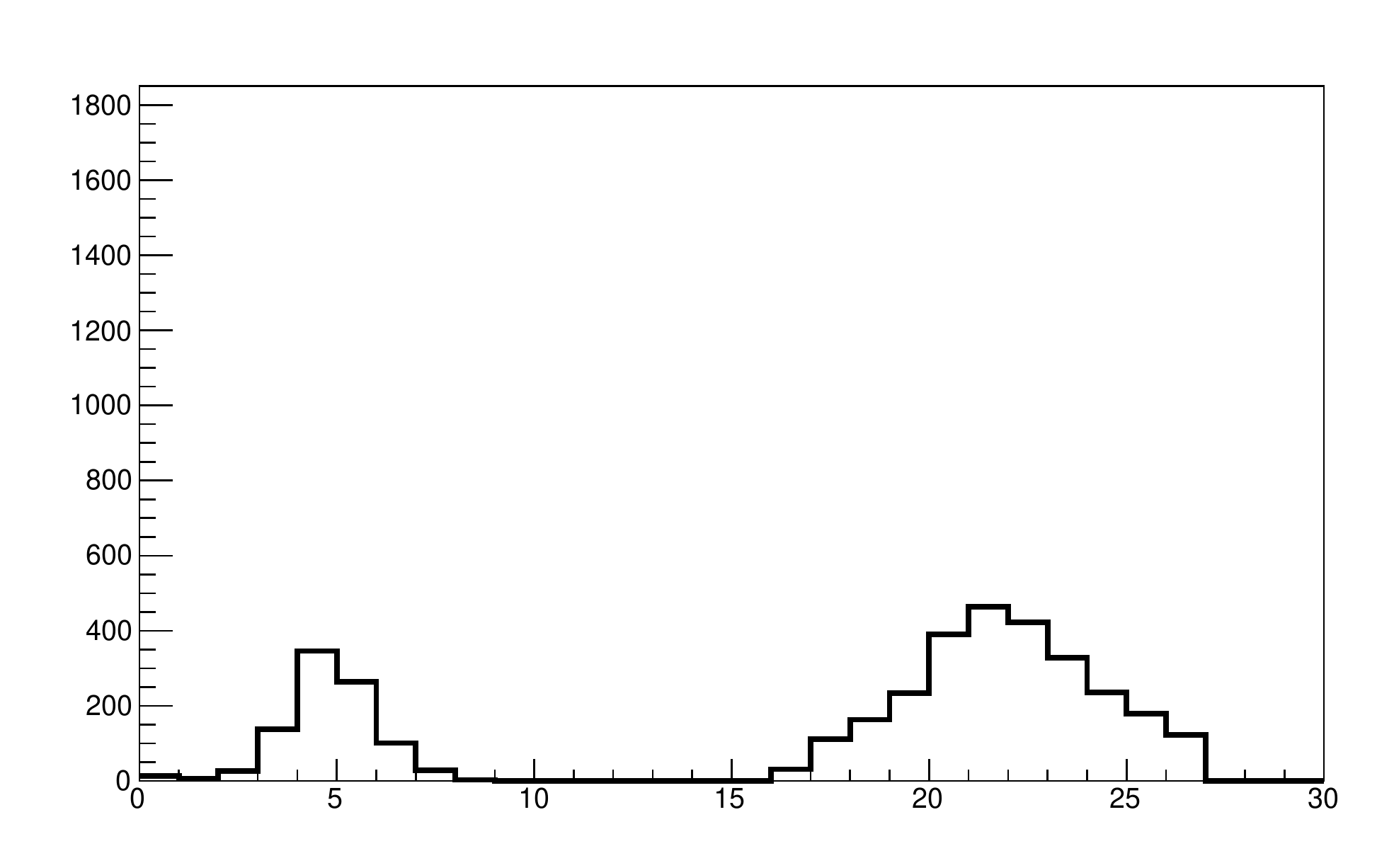}
}
\subfloat[strip 2\label{fig: time_structure_2}]{
\includegraphics[width=.32\textwidth]{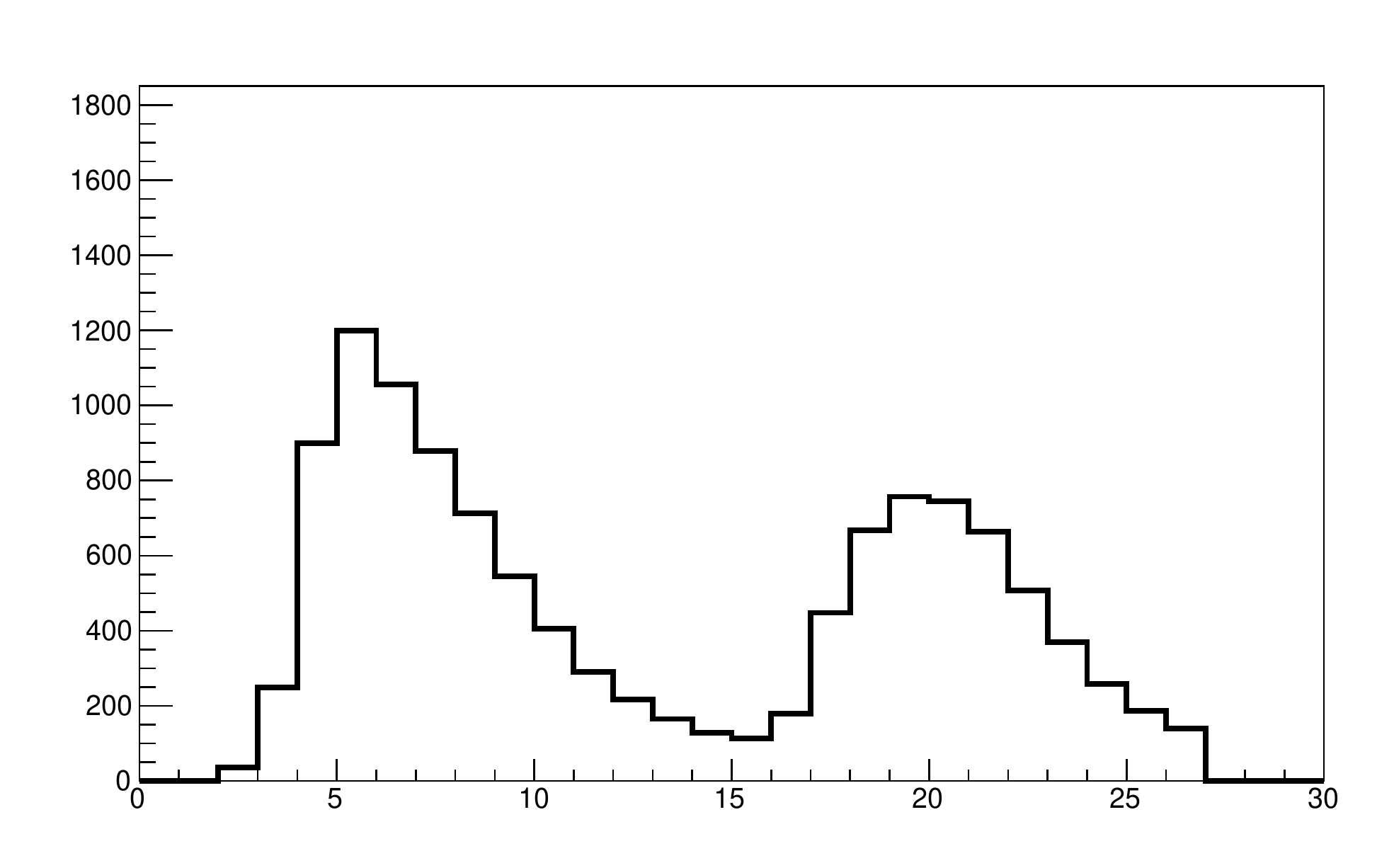}
}
\subfloat[strip 3\label{fig: time_structure_3}]{
\includegraphics[width=.32\textwidth]{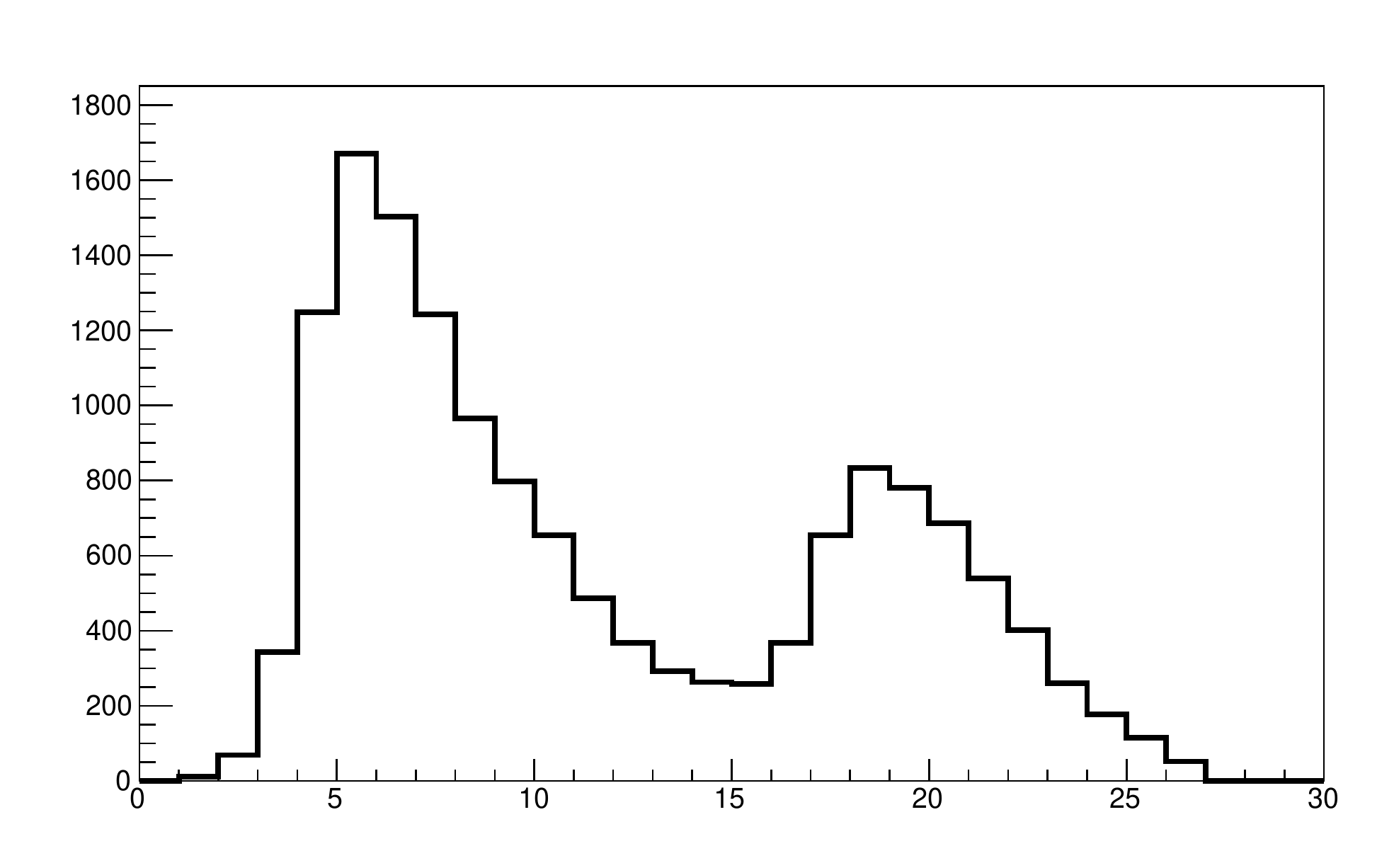}
}
\\
\subfloat[strip 4\label{fig: time_structure_4}]{
\includegraphics[width=.32\textwidth]{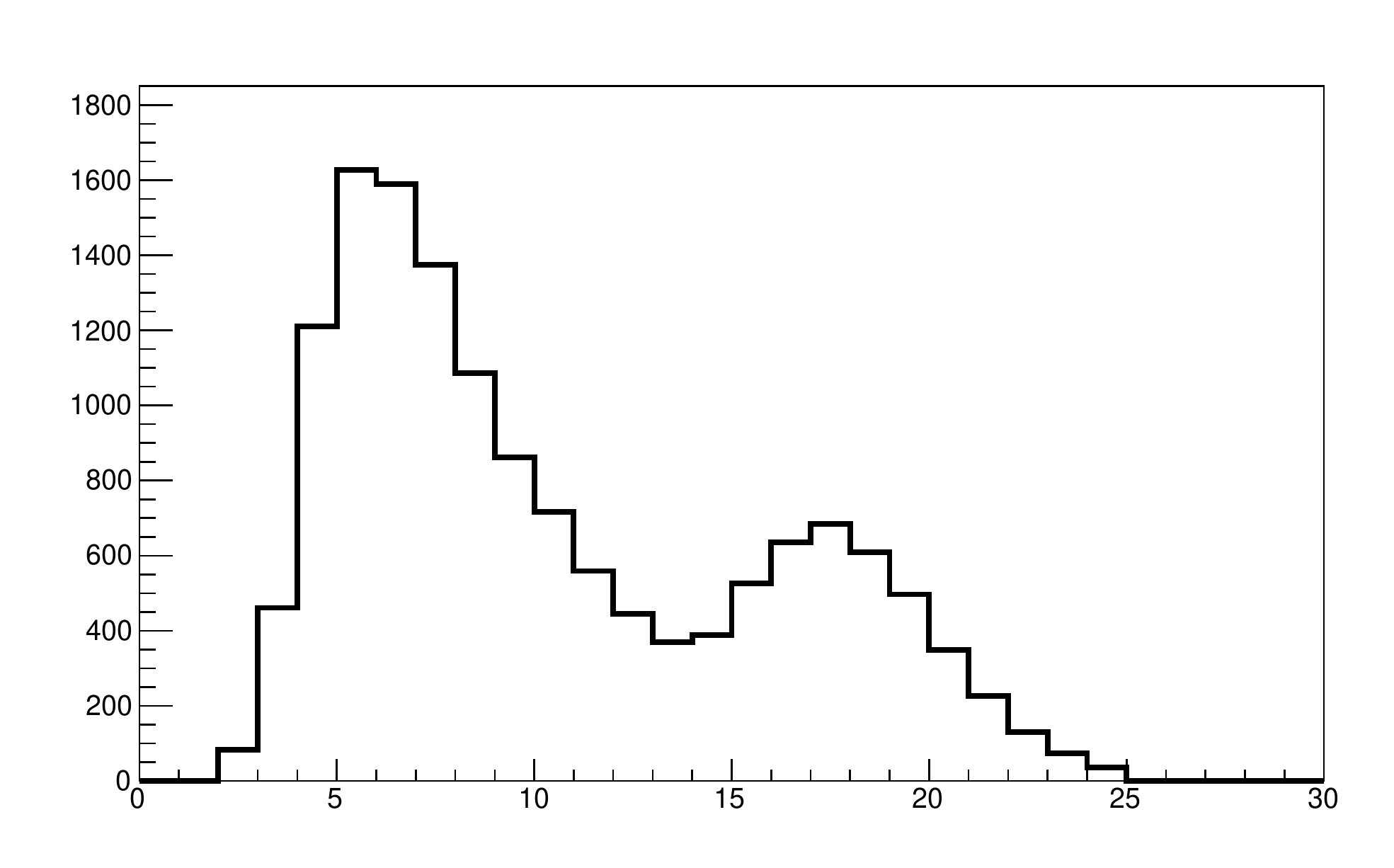}
}
\subfloat[strip 5\label{fig: time_structure_5}]{
\includegraphics[width=.32\textwidth]{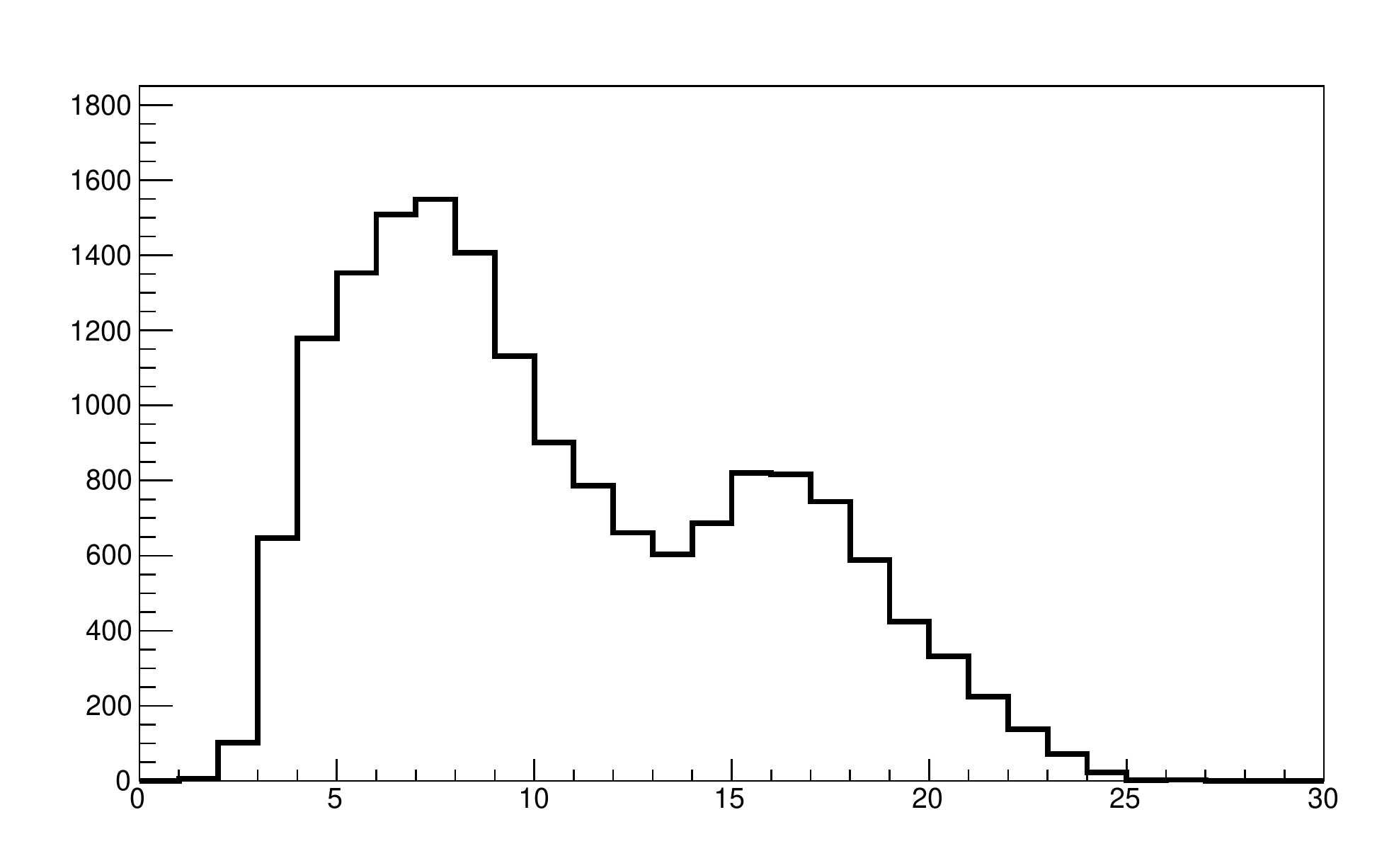}
}
\subfloat[strip 6\label{fig: time_structure_6}]{
\includegraphics[width=.32\textwidth]{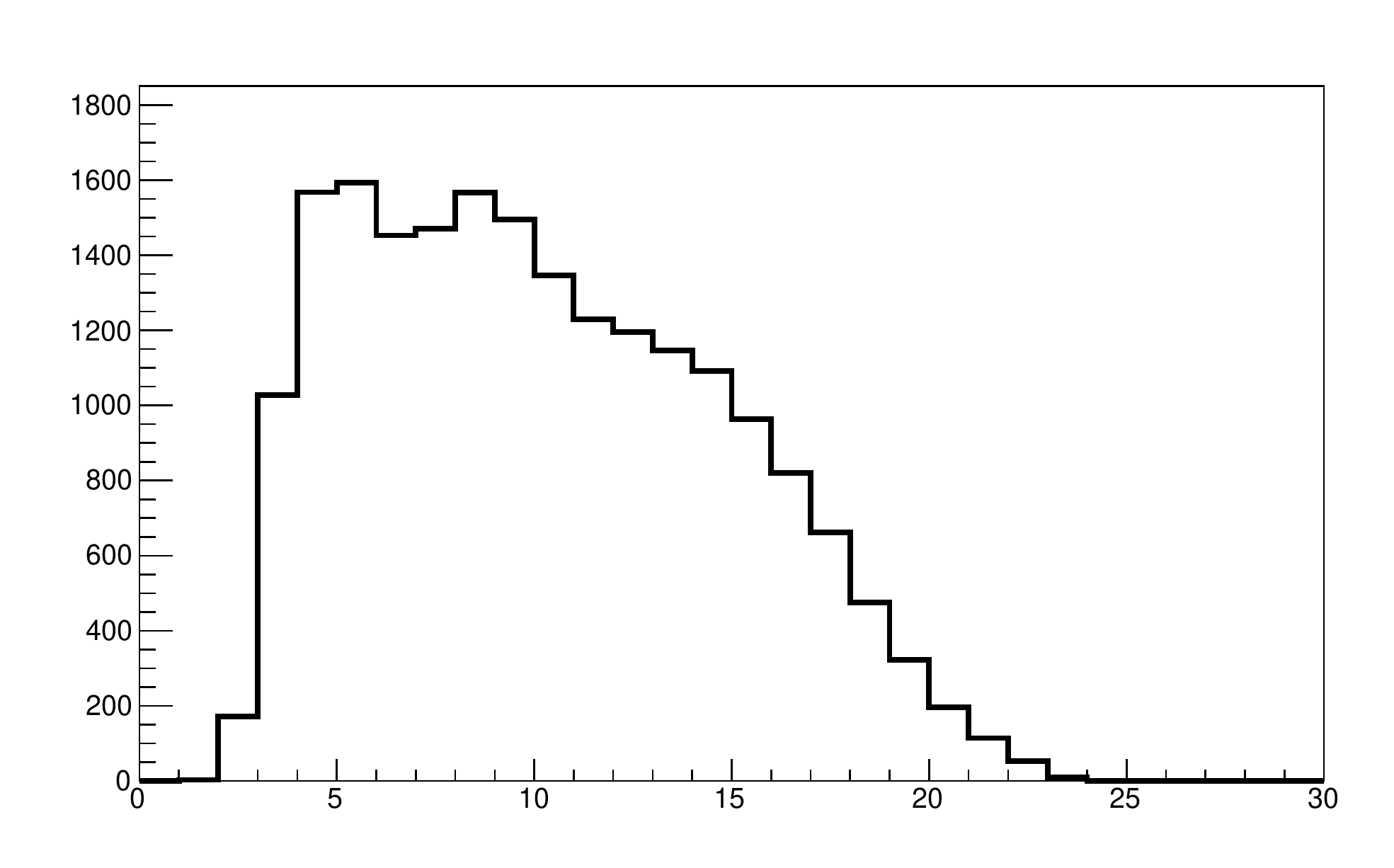}
}
\\
\subfloat[strip 7\label{fig: time_structure_7}]{
\includegraphics[width=.32\textwidth]{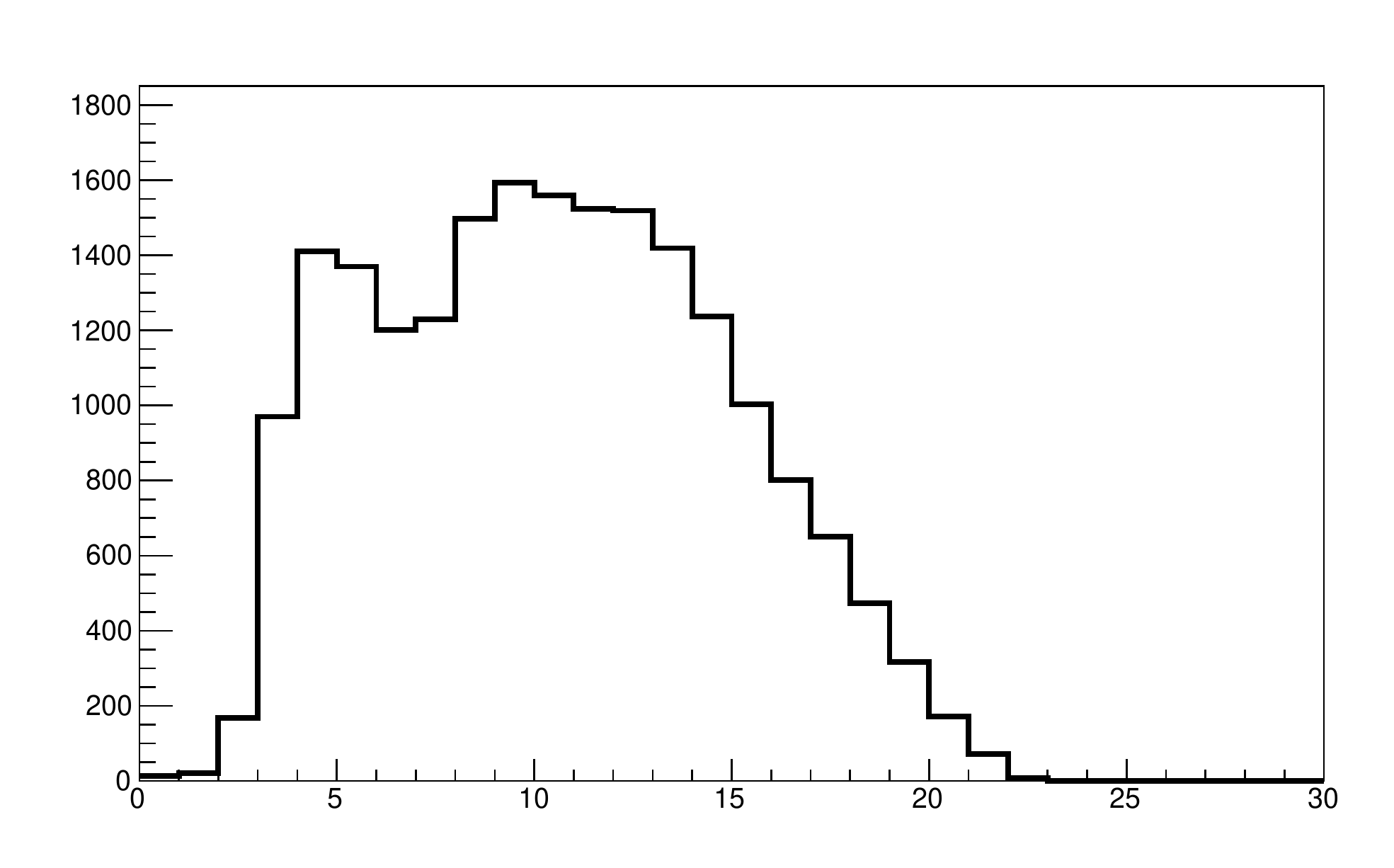}
}
\subfloat[strip 8\label{fig: time_structure_8}]{
\includegraphics[width=.32\textwidth]{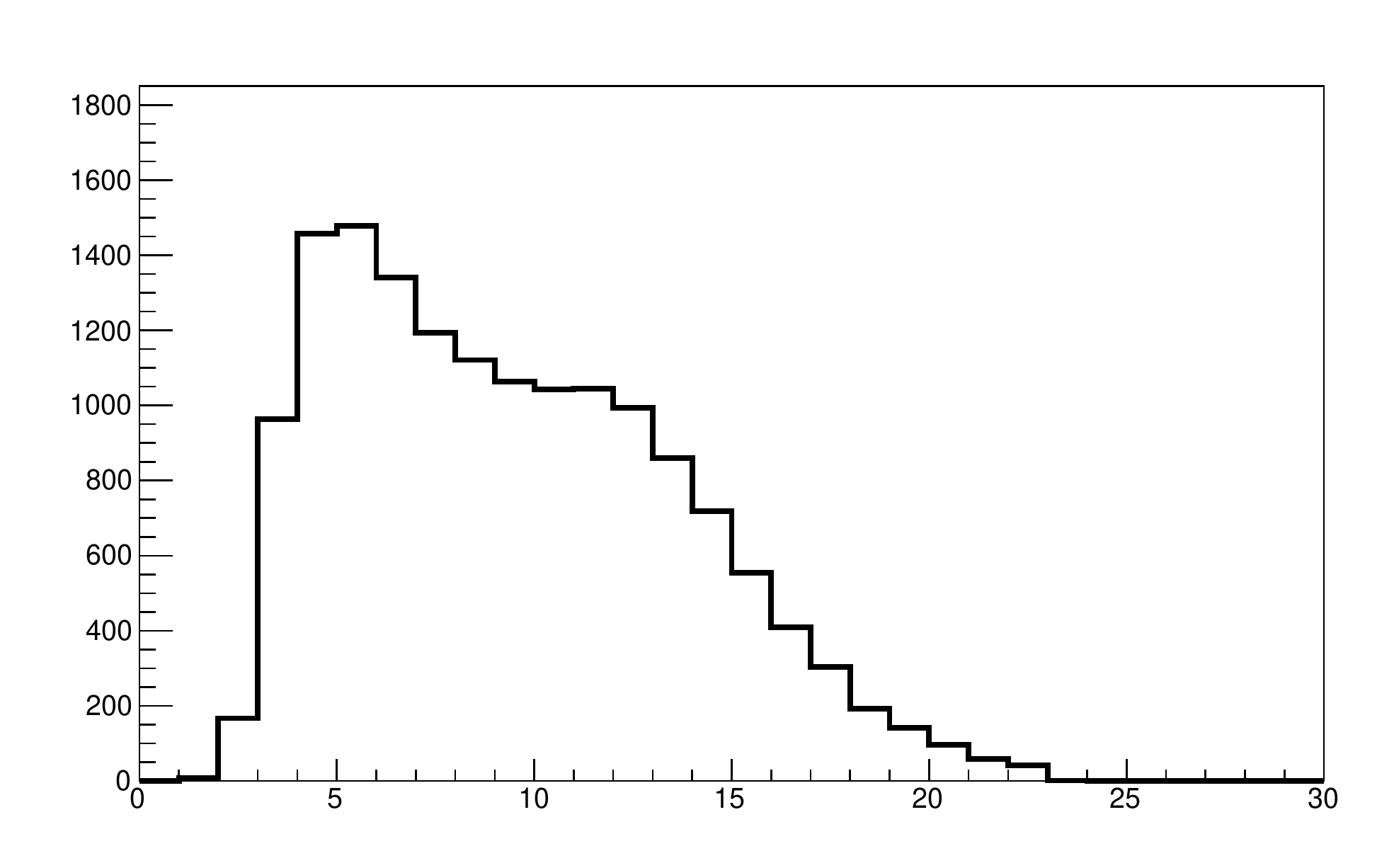}
}
\subfloat[strip 9\label{fig: time_structure_9}]{
\includegraphics[width=.32\textwidth]{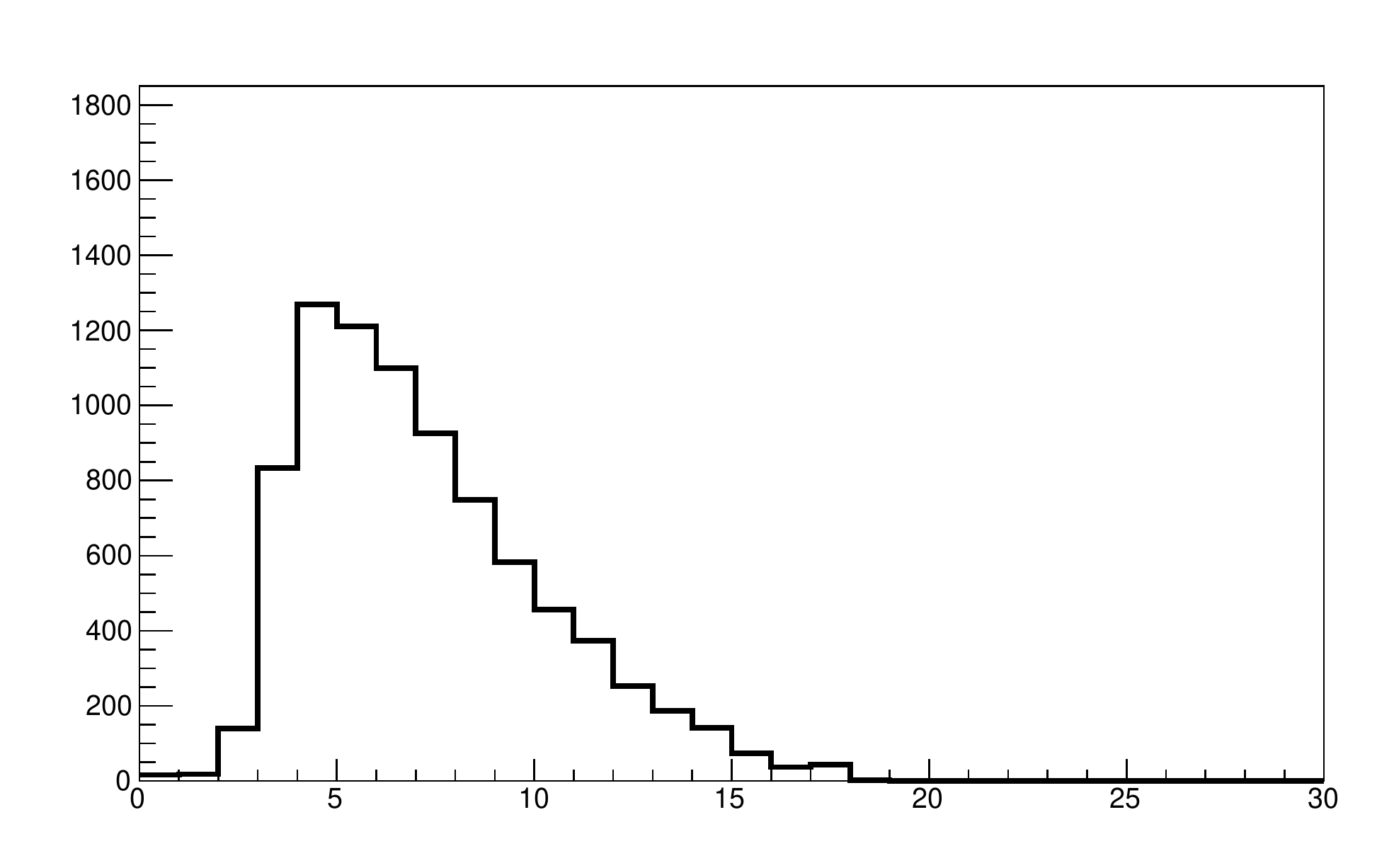}
}
\caption{Typical time structure of the neutron conversion signal in different strips with the 25 ns time bins on the x-axis, and the charge in arbitrary units on the y-axis.}
\label{fig: time_structure}
\end{figure}

The typical signals due to a neutron conversion are shown in figure~\ref{fig: time_structure}. The figure depicts the raw waveforms from the y view on different strips, digitized every 25~ns. The signals on different strips are not synchronous because different drift depths require different drift times. During the drift in electric fields, electrons diffuse as a result of multiple collisions with the gas molecules, and the initially localized charge cloud spreads out. The diffusion depends on the gas mixture, the pressure and the electrical field. At normal temperature and pressure (NTP) in Ar/CO$_2$ 70/30 at a field of 700V/cm, the $\sigma$ of the transversal and longitudinal diffusion amounted to around 160~$\mu$m over the 10 mm drift~\cite{Gas_detectors}. An ionizing particle crossing a portion of the active volume creates electrons and ions that move in opposite directions separated by the electric field. The speed of the conversion electrons is much larger than the electron drift velocity, therefore the primary charge can be assumed to be released instantaneously along their path. The electron cloud moves rigidly at a constant speed due to the uniform electric field. Neglecting the small amount of diffusion, the electron cloud preserves its original track shape along the drift.

\begin{figure}[htbp]
\centering
\subfloat[x view\label{fig: track_Gd_x}]{
\includegraphics[width=.49\textwidth]{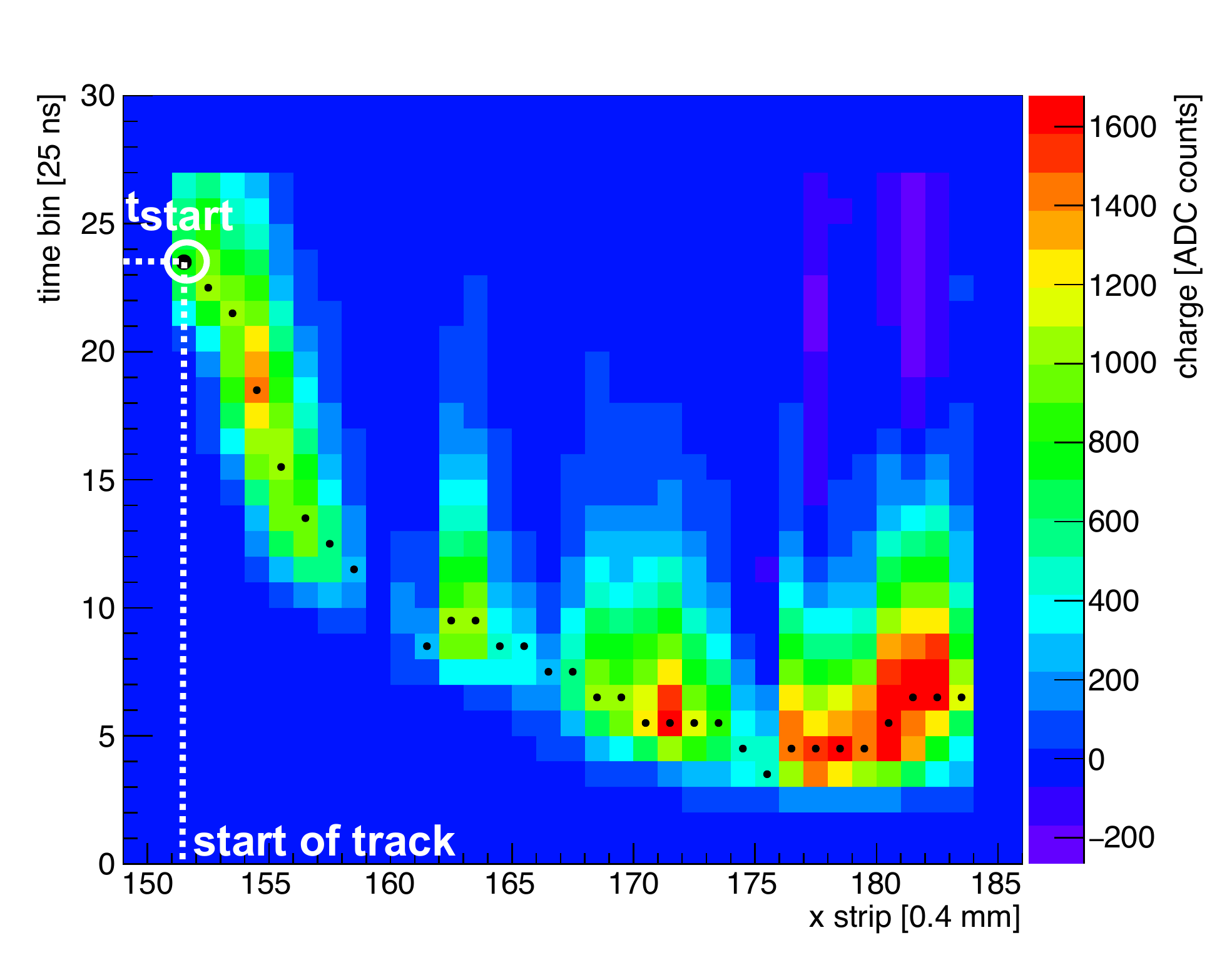}%
}
\subfloat[y view\label{fig: track_Gd_y}]{
\includegraphics[width=.49\textwidth]{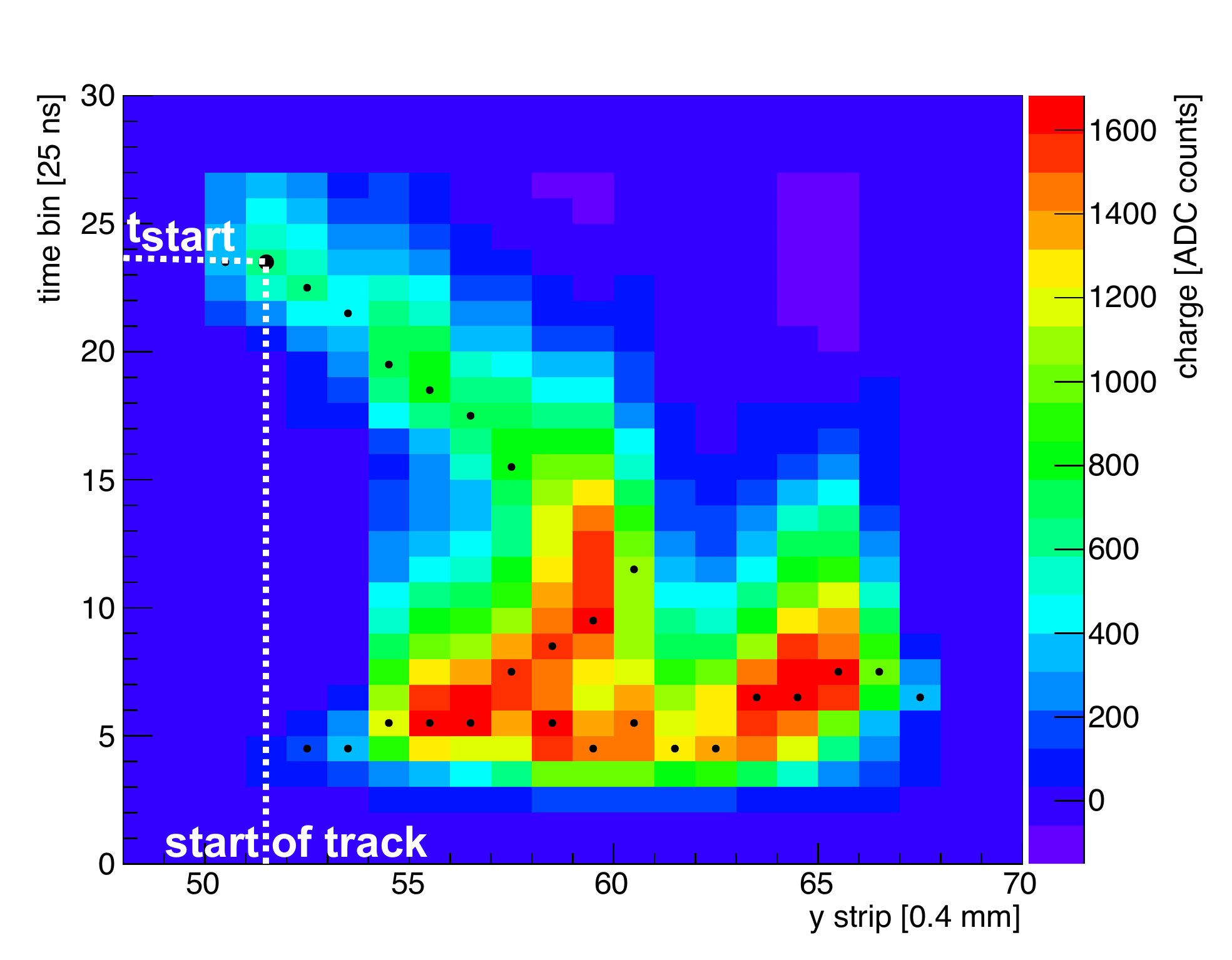}%
}
\caption{Typical electron track in the drift volume of the Triple-GEM detector with Gd cathode.}
\label{fig: track_Gd_gem}
\end{figure}

Knowing the instant when the primary charge is released ($t_0$), one can extract the drift time and consequently the absolute position along the drift of the ionizing event. In the present configuration this information is not accessible. Nevertheless, in the case of straight tracks from $\alpha$ particle or muons, the electrons from the end of the track always induce the signal before the electrons from the beginning of the track, unless the particle moves in a direction parallel to the cathode. This concept forms the base of the Time Projection Chamber (TPC)~\cite{TPC}, which allows the identification of the beginning of the track even without knowing $t_0$. Conversion electrons on the other hand quite frequently propagate in a different fashion. They first move away from their origin in the converter on the cathode, but can later turn and move back towards the cathode. The x and y view of such a typical curved track as measured with the Gd-GEM setup is depicted in figure~\ref{fig: track_Gd_gem}. Obviously the signal from the end of the conversion electron track does not always arrive first. But it still holds true that the electrons from the beginning of the track have the largest drift time. The modified $\mu$TPC analysis for gadolinium is thus only able to determine the start of a track.

\begin{figure}[htbp]
\centering
\subfloat[Discriminator rising flank\label{fig: rising flank}]{
\includegraphics[width=.48\textwidth]{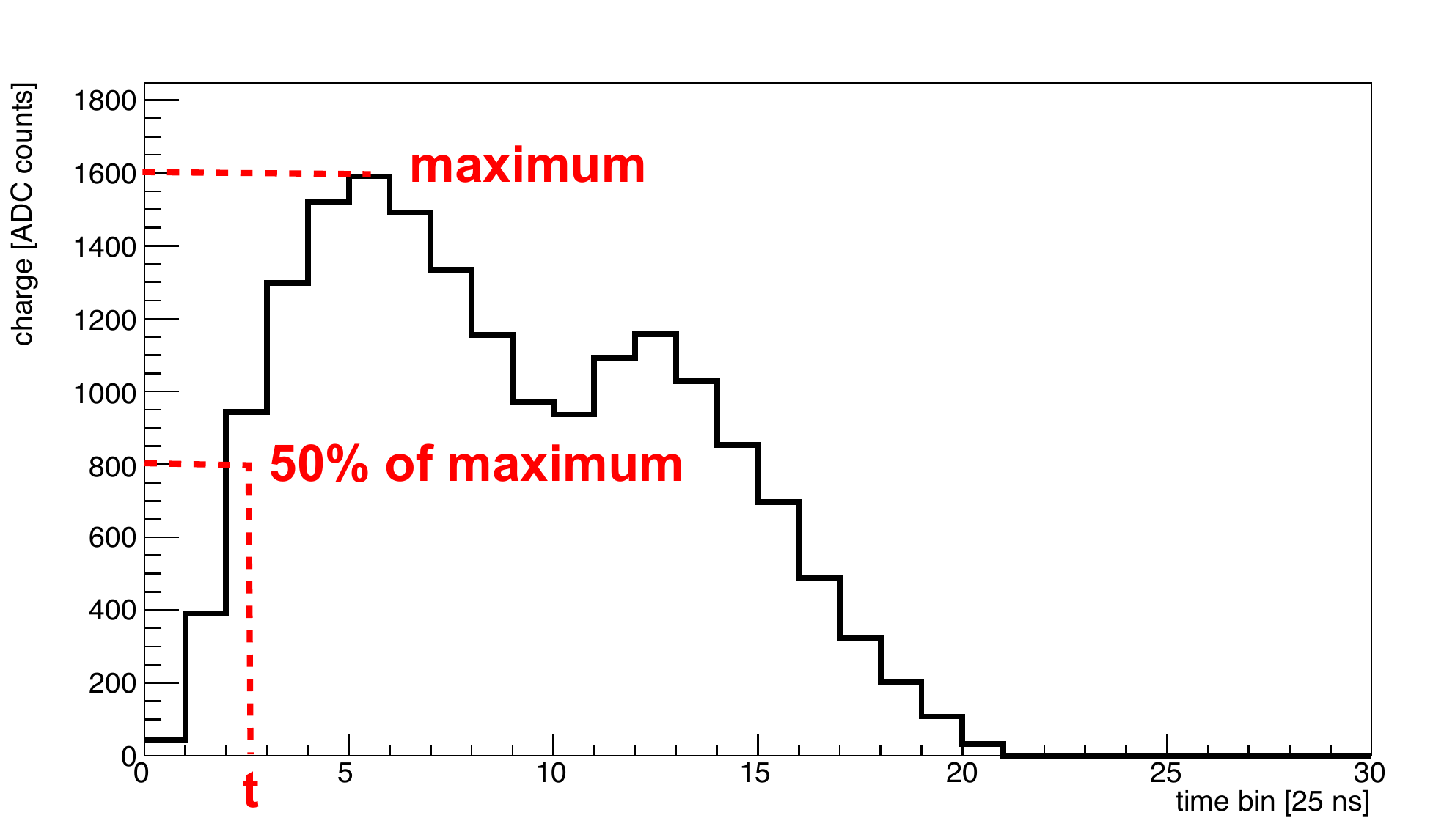}
}
\subfloat[Global maximum\label{fig: global maximum}]{
\includegraphics[width=.48\textwidth]{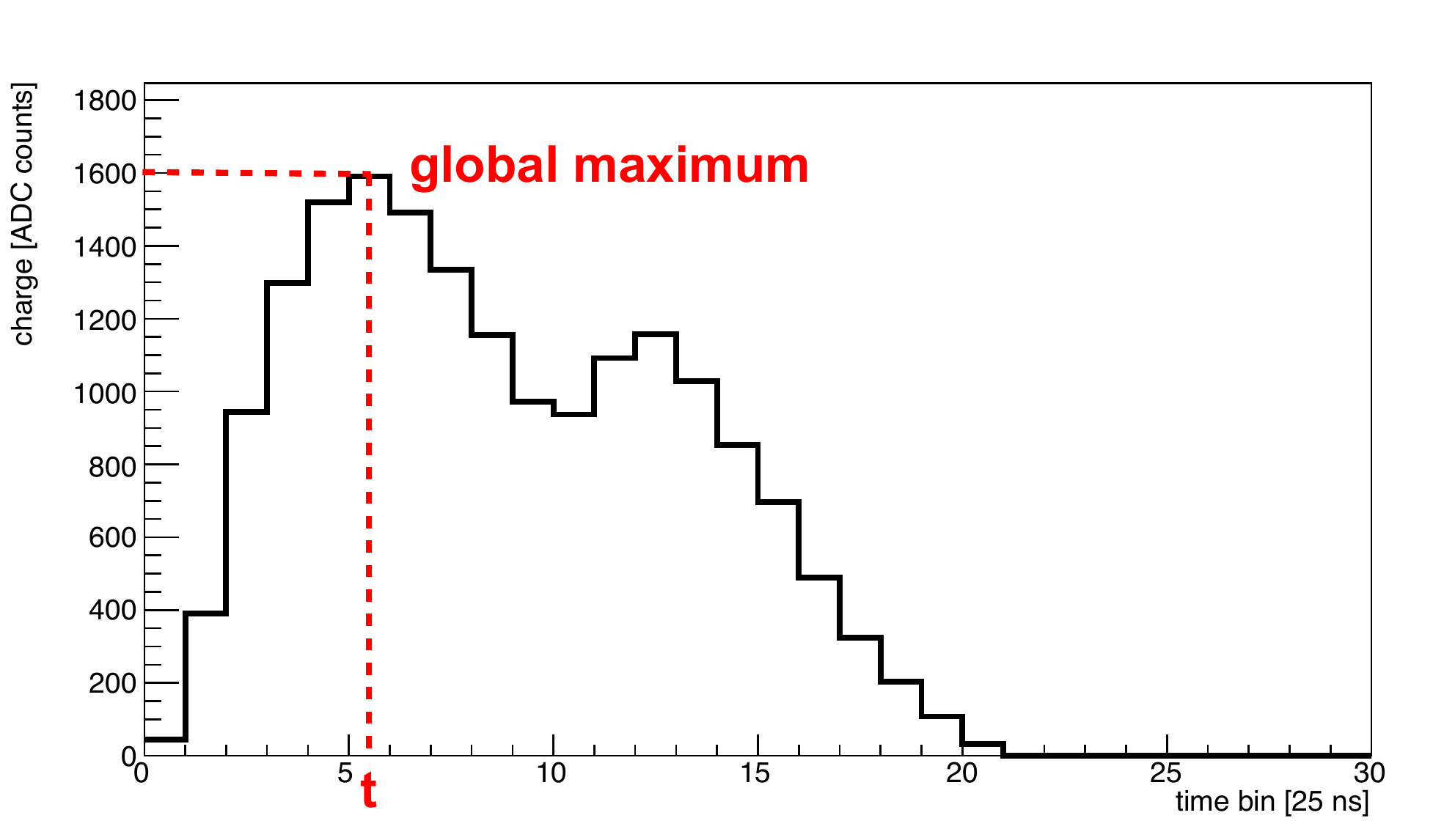}
}
\\
\subfloat[Discriminator falling flank\label{fig: falling flank}]{
\includegraphics[width=.48\textwidth]{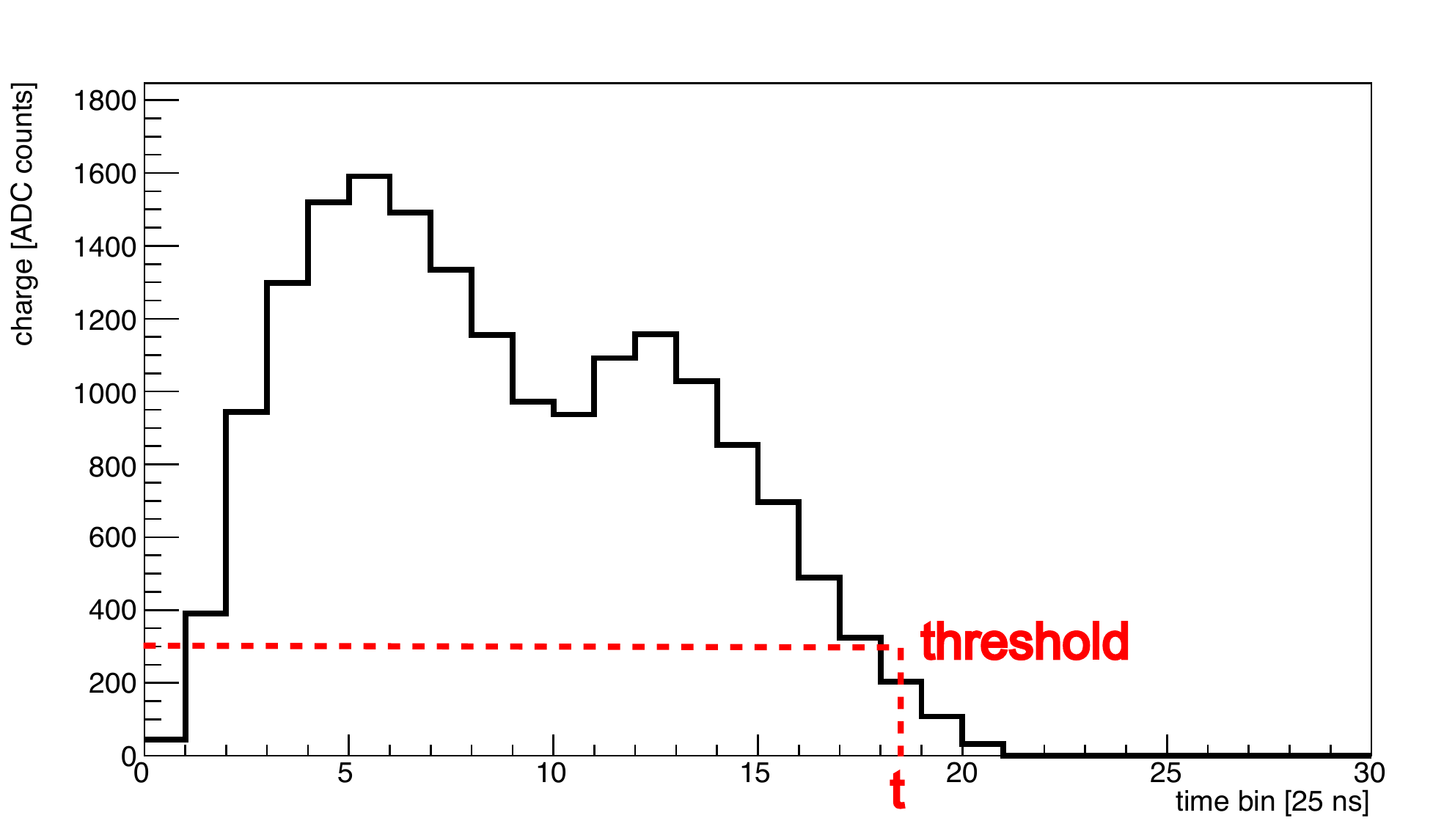}
}
\subfloat[Last local maximum\label{fig: last local maximum}]{
\includegraphics[width=.48\textwidth]{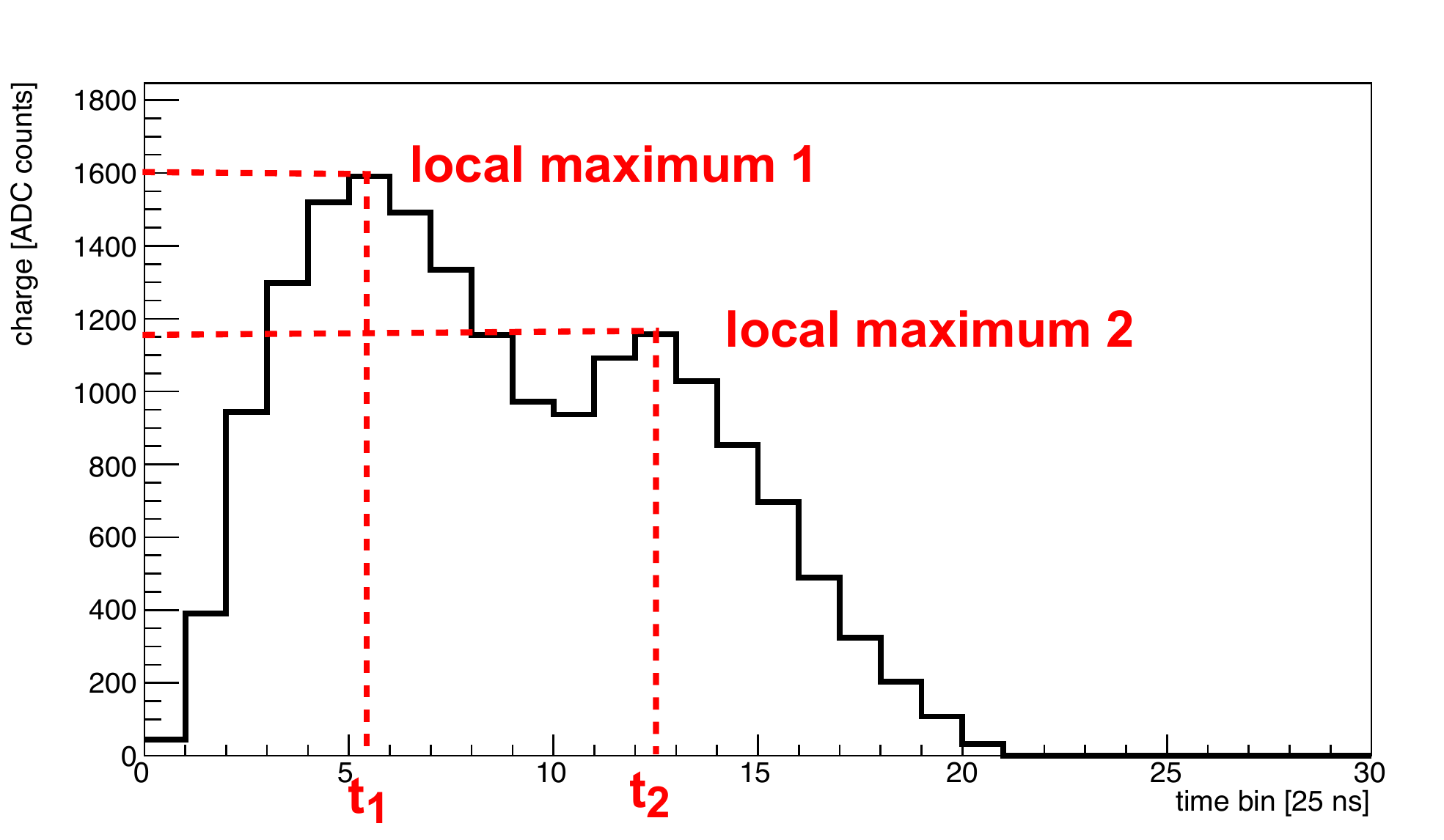}
}
\caption{Different algorithms to determine the strip of origin of the conversion electron.}
\label{fig: algorithms}
\end{figure}

As shown in figure~\ref{fig: time_structure}, the waveform of the signal from the gadolinium conversion electrons contains quite frequently several local maxima. This fact stems from the changes of direction and the resulting curved path that the conversion electrons travel in the drift volume. However with the $\alpha$ particles from $^{10}$B$_4$C~\cite{uTPC_Boron} and their straight path, the waveforms just had one maximum. To identify the start of the track, it was thus sufficient to determine for each strip the time \emph{t} where the rising flank of the pulse reaches 50$\%$ of its maximum amplitude (\ref{fig: rising flank}). The start of the track was then the strip with the largest \emph{t}. For the Gd conversion electrons one reaches better results when using the time of the global maximum (\ref{fig: global maximum}), the time when the falling flank falls below a set threshold (\ref{fig: falling flank}) or the time of the last local maximum (\ref{fig: last local maximum}). Interpreting the track as a point cloud composed of local maxima, an approach based on the Principal Component Analysis (PCA)~\cite{ParametricCurves} has been tested to reconstruct the whole track. It did not lead to an improvement in position resolution compared to the last-local-maximum algorithm, and was discarded due to the considerable computational effort.

\begin{figure}[htbp]
\centering
\subfloat[250 $\mu$m thick Gd cathode\label{fig: cathode}]{
\includegraphics[width=.46\textwidth]{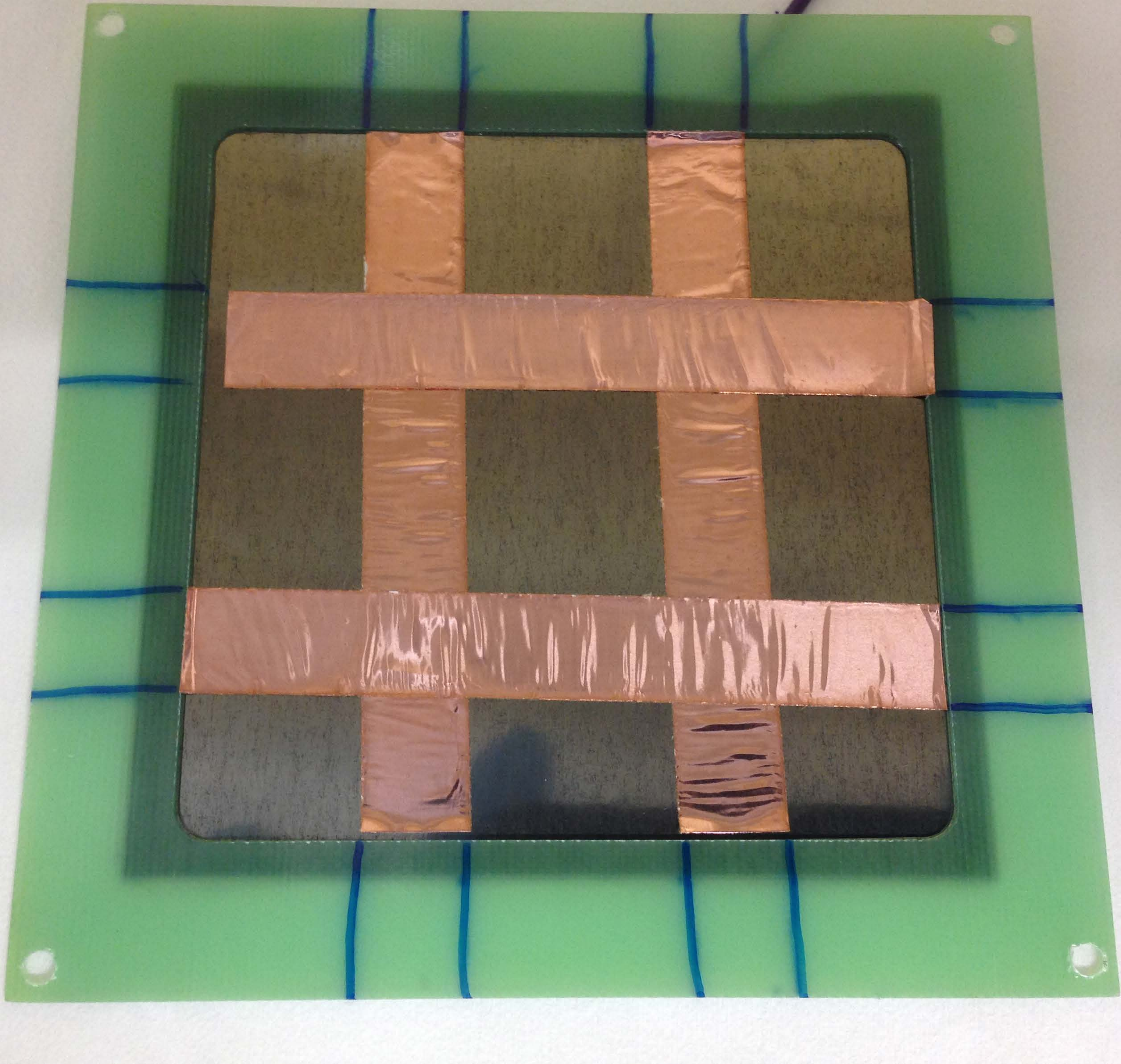}%
}
\subfloat[Hit distribution 0.3 x 10~cm$^{2}$\label{fig: hit_distribution_3mm_10cm}]{
\includegraphics[width=.52\textwidth]{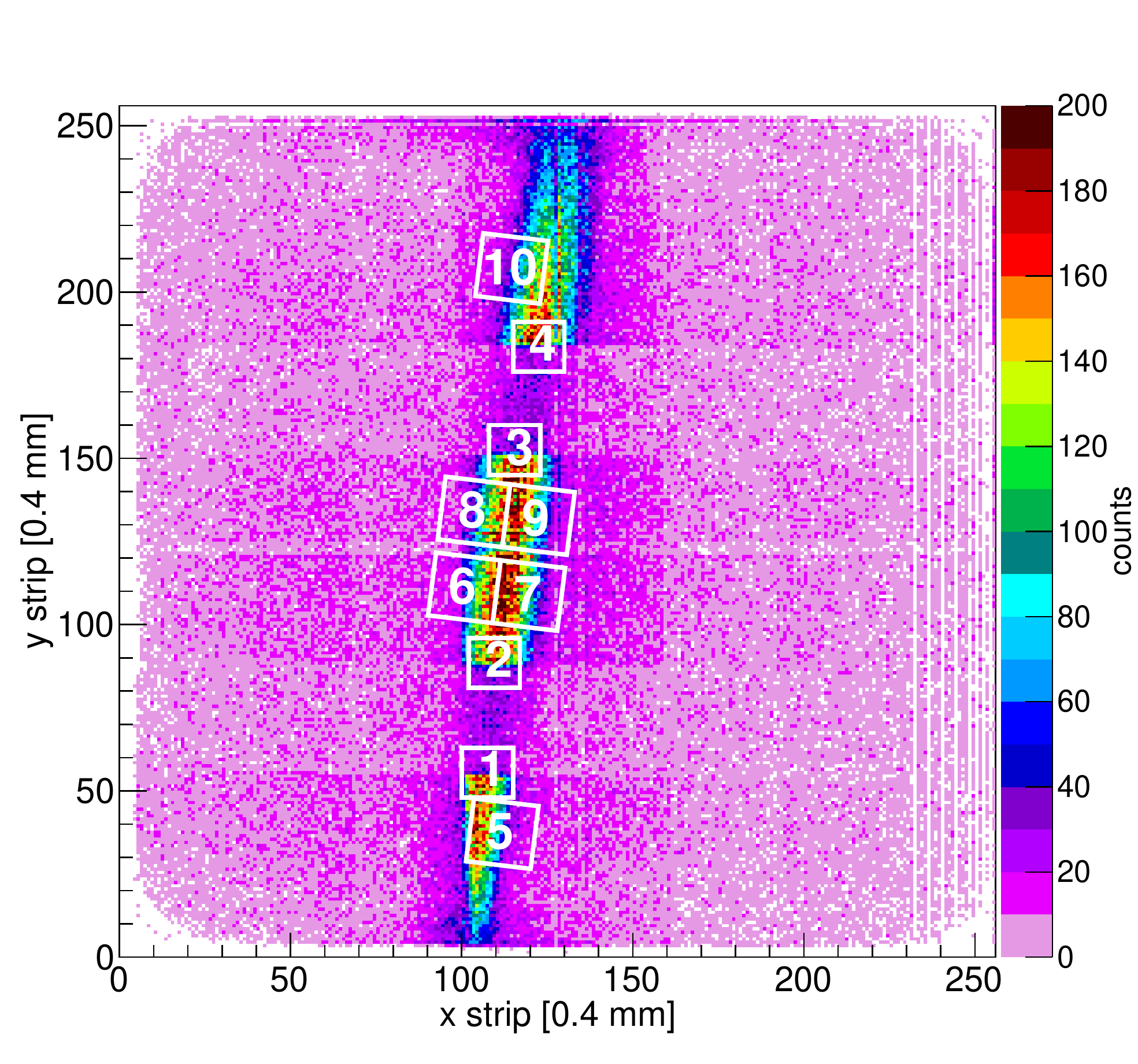}%
}
\caption{Gd cathode with copper tape and resulting hit distribution (reconstructed using the last-local-maximum) with regions of interest.}
\label{fig: single_gem_hits}
\end{figure}

Figure~\ref{fig: cathode} shows the cathode with copper grid that was used for the measurement. The copper grid was added to stop the conversion electrons from escaping from the converter into the gas. It provides a sharp edge to measure the position resolution in addition to the edge of the collimated beam, which is widened by the scattering of the neutrons in readout board and GEM foils. Both types of edges can be seen in the plot of the hit distribution~\ref{fig: hit_distribution_3mm_10cm}, obtained by the reconstruction of the hit position using the last-local-maximum algorithm. To determine the improvement in position resolution due to the $\mu$TPC analysis compared to the traditional centroid-of-charge approach, the regions with a large step in the hit distribution were studied. 

\begin{figure}[htbp]
\centering
\includegraphics[width=0.95\textwidth]{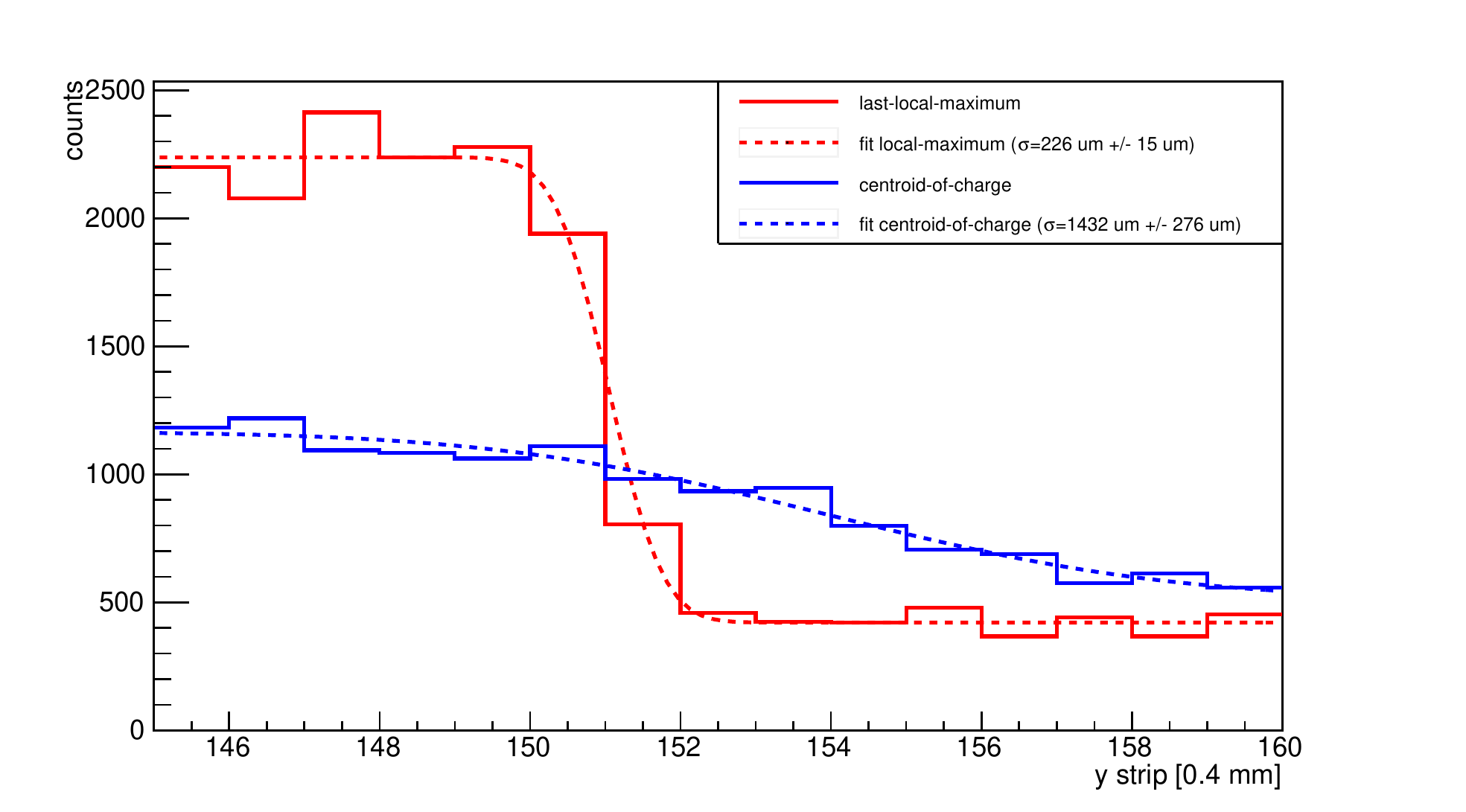}
\caption{Region R3 (0.3 x 10 cm$^{2}$ collimated beam): Distribution of the reconstructed y coordinate using a centroid-of-charge-based technique (in blue) and the $\mu$TPC analysis (in red). }
\centering
\label{fig: fit_y}
\end{figure}

Figure~\ref{fig: fit_y} shows as example the analysis of the position resolution in region \emph{R3} of the 0.3 x 10~cm$^{2}$ collimated beam in figure~\ref{fig: hit_distribution_3mm_10cm}. The position resolution (standard deviation $\sigma$) is extracted from the fit of the complementary error function to the data. The reconstructed position obtained with the $\mu$TPC concept (last-local-maximum) is shown in red, the centroid-of-charge approach in blue. For the conversion electrons from Gd, the centroid-of-charge approach cannot be seriously considered as solution to determine the position resolution. It is thus shown here for comparison purposes only. For the regions with copper tape and the beam regions, last-local-maximum, falling-flank and global-maximum algorithm lead to similar results within errors. Intuitively the falling-flank algorithm should lead to the best results, since the falling flank has always a closer proximity to the start of the track than the global or last local maximum. But depending on the amplifier noise levels, the shaping curve and the chosen threshold, the results may vary considerably. The last local or global maximum on the other hand can be determined with less ambiguity. To evaluate the particular differences between these three algorithms, more studies with a larger number of events are needed. The influence of the topology of an individual track on the results has to be reduced.

The results of all analysed regions can be found in table~\ref{table: results_3mm_10cm}. For the tape edges, the $\sigma$ of the position resolution for the $\mu$TPC technique is on average better than 250~$\mu$m.  and 1.7~mm or better for the beam edges. The intrinsic position resolution seems to be closely related to the 400~$\mu$m strip pitch of the x/y readout. The values for the beam edges are indicative of the divergence of the collimation system and the scattering of the beam. A $\sigma$ of about 1.2~mm has been determined with a Monte Carlo simulation for the divergence of the beam. For the scattering of the beam in the detector under test, a $\sigma$ of 124~$\mu$m has been obtained with Geant4. The position resolution measured with the edge of the beam is thus clearly dominated by the divergence of the neutron beam. 

\begin{table}
\centering
\footnotesize
\begin{threeparttable}
\begin{tabular}[htbp]{ l|r|r|r|r|}
\cline{2-5}
& \multicolumn{1}{ |l| }{last-local-maximum} & \multicolumn{1}{ |c| }{falling-flank} & \multicolumn{1}{ |l| }{global- maximum} & \multicolumn{1}{ |c| }{centroid-of-charge} \\ 
\hline
\multicolumn{1}{ |l|}{Position} & $\sigma$ [$\mu$m] & $\sigma$ [$\mu$m] & $\sigma$ [$\mu$m] & $\sigma$ [$\mu$m]  \\
\hline
\multicolumn{1}{ |l|}{tape R1} & 175(14) & 162(22) & 188(20) & 627(151)  \\ 
\multicolumn{1}{ |l|}{tape R2} & 241(16) & 308(21) & 346(30) & 2217(608) \\ 
\multicolumn{1}{ |l|}{tape R3} & 226(15) & 262(20) & 221(18) & 1432(276) \\ 
\multicolumn{1}{ |l|}{tape R4} & 171(12) & 117(61) & 204(21) & 1071(208) \\ 
\multicolumn{1}{ |l|}{beam R5} & 1090(33) & 955(29) & 1146(37) & 3325(638)\\ 
\multicolumn{1}{ |l|}{beam R6} & 1378(41) & 1182(36) & 1306(43) & 3152(606)\\ 
\multicolumn{1}{ |l|}{beam R7} & 1452(39) & 1281(34) & 1322(37) & 1401(94)\\ 
\multicolumn{1}{ |l|}{beam R8} & 1635(51) & 1459(44) & 1479(52) & 2884(559)\\ 
\multicolumn{1}{ |l|}{beam R9} & 1360(38) & 1235(35) & 1251(39) & 2918(141)\\ 
\multicolumn{1}{ |l|}{beam R10} & 1709(71) & 1433(54) & 1692(79) & - \tnote{*}\\  
\hline
\multicolumn{1}{ |l|}{weighted mean tape } & 198(17) & 243(36) & 223(28) & 931(222) \\ 
\hline
\multicolumn{1}{ |l|}{weighted mean beam } & 1358(86) & 1202(76) & 1302(59) & 1927(366) \\ 
\hline

\end{tabular}
\caption{Fit results for the 0.3 x 10 cm$^{2}$ collimated beam and the regions of interest depicted in figure~\protect\ref{fig: hit_distribution_3mm_10cm}.}
\label{table: results_3mm_10cm}
\begin{tablenotes}
\item[*] Fit not possible (edge very weak in data).
\end{tablenotes}
\end{threeparttable}
\end{table}

\FloatBarrier
\newpage

\section{Conclusions and Outlook}
\label{sec:conclusions}
After already having demonstrated the $\mu$TPC concept with thermal neutrons using $^{10}$B$_4$C as neutron converter~\cite{uTPC_Boron}, it is here for the first time applied to data from a Gd-based neutron detector. The results show that the $\sigma$ of the intrinsic position resolution is on average better than 250~$\mu$m. The achievable position resolution seems to be about half the pitch of the x/y strip readout. Considering the necessary detector gain, it is possible to use only two GEM foils to reduce the amount of Kapton in the detector. Together with the development of a readout made from material with a lower scattering cross section for cold neutrons, this will improve the detection efficiency and reduce the impact of the scattering on position resolution. To determine the position resolution without influence of the divergence of the beam, measurements with a cadmium or gadolinium mask in front of the readout board are planned. To improve the detector efficiency, either the addition of a second detector in forwards direction or the use of enriched Gd converters has to be studied.

\section{Acknowledgements}
\label{sec:Acknowledgements}
This work was partially funded by the EU Horizon 2020 framework, BrightnESS project 676548.

\FloatBarrier

\end{document}